\documentclass[a4paper,11pt]{article}
\usepackage{jheppub} 
\usepackage{lineno}

\usepackage{amsmath}
\usepackage{graphicx}
\usepackage{mathrsfs}
\usepackage{hyperref}
\usepackage{xcolor}
\usepackage{subfigure}
\usepackage{slashed}
\usepackage{braket}
\usepackage{epsfig,color}
\usepackage{multirow}
\usepackage{commath}
\usepackage{adjustbox}

\usepackage{amssymb}

\usepackage{float}

\newcommand{\MSbar}{\overline{\text{MS}}}
\newcommand{\muMS}{\mu_{\MSbar}}
\newcommand{\muRI}{\mu_{\rm RI}}

\begin{document}

\title{Two-loop RI/SMOM-$\overline{\text{MS}}$ matching factors for the renormalization of bi-local operators contributing to $\epsilon_K$}

\newcommand\cu{Physics Department, Columbia University, New York, NY 10027, USA}

\author{Erik Lundstrum}
\emailAdd{ell2156@columbia.edu}
\affiliation{\cu}

\abstract{
    The measure of indirect CP violation in the neutral kaon system, $\epsilon_K$, is an important precision test of the Standard Model. A first-principles Standard Model prediction of $\epsilon_K$ requires evaluating the long-distance contribution in which the $\Delta S = 2$ process proceeds via two $\Delta S = 1$ operators. The non-perturbative nature of QCD in the relevant energy regimes requires a lattice QCD simulation. In this paper we present the two-loop, $O(\alpha_s)$ RI/SMOM - $\overline{\text{MS}}$ matching factors required to renormalize a lattice QCD calculation of the long-distance contribution to $\epsilon_K$.
}

\maketitle

\section{Introduction}
The neutral kaon system is an important setting for precision tests of the Standard Model (SM). Processes like CP-violating $K \rightarrow \pi\pi$ decays and $\Delta M_K$ in kaon mixing are highly suppressed in the SM and are therefore promising areas in which to observe beyond the Standard Model physics at work. Furthermore, the parameters $\Delta M_K$, $\epsilon_K$ and $\varepsilon'$ have been precisely measured experimentally and therefore precise, first-principles SM predictions are of immediate interest. In this work we focus on the parameter $\epsilon_K$ which describes indirect CP violation in $K\rightarrow \pi\pi$ decays. Because $\epsilon_K$ is a CP-violating process, the relevant CKM matrix elements don't suppress the top quark contribution. Therefore, $\epsilon_K$ is dominated by a short-distance contribution described by a single, local $\Delta S = 2$ operator. There is, however, a subdominant ($\sim 5\%$) but phenomenologically important contribution coming from lower energy scales in which the $\Delta S = 2$ process proceeds via two $\Delta S = 1$ operators separated by a larger distance and joined by a closed quark loop. The presence of the top quark breaks the Glashow-Iliopoulos-Maiani (GIM) cancellation in this closed quark loop, resulting in a short-distance, logarithmic divergence when the positions of the two $\Delta S = 1$ operators collide.  

A first principles SM prediction of this long-distance contribution with controlled systematic errors is at present only possible using a lattice QCD simulation. Such simulations have recently been performed by the RBC/UKQCD collaboration. An exploratory, proof-of-concept calculation was published in Ref.~\cite{Bai:2023lkr} with unphysical quark masses. More recently, a lattice QCD simulation using physical quark masses is being performed and will soon appear. For an in-progress summary of the efforts at physical quark masses, see Refs.~\cite{Huo:2025bhq, Yikai_thesis}. In these lattice QCD calculations, the W boson and the top and bottom quarks are integrated out while the charm is left active. We are left with an effective four-quark realization of the weak interactions. The operator product expansion is used to determine the SM prediction by combining matrix elements of the four-quark operators evaluated on the lattice with a set of Wilson coefficients which encode the short-distance effects. The latter are calculated perturbatively in QCD + electroweak perturbation theory in the $\overline{\text{MS}}$ renormalization scheme, see Ref.~\cite{Buchalla:1995vs} for a detailed review. In order to combine the lattice-regulated matrix elements with the $\overline{\text{MS}}$ Wilson coefficients, one uses the regularization-independent (RI) renormalization scheme introduced in Ref.~\cite{Martinelli:1994ty} as an intermediate scheme. The factors which relate the RI scheme to the $\overline{\text{MS}}$ scheme are calculated in QCD perturbation theory, while the factors which convert between the lattice-regulated and RI schemes are determined non-perturbatively on the lattice. The lattice QCD calculations of the long-distance contribution to $\epsilon_K$ in Refs.~\cite{Bai:2023lkr, Huo:2025bhq, Yikai_thesis} have suffered from an incomplete knowledge of the RI-$\overline{\text{MS}}$ matching factors for the bi-local operator at $O(\alpha_s)$. Because this process proceeds via two $\Delta S = 1$ operators, the one-loop contribution comes from a closed quark loop and the $O(\alpha_s)$ contributions only arise at two-loop order in perturbation theory. It is the purpose of this paper to calculate these matching factors to $O(\alpha_s)$ by completing the required two-loop calculation. 

An interesting aspect of this paper which we would like to emphasize is that all of the code required for this calculation was produced by prompting the agentic AI system Claude Fable 5.1 \cite{anthropic2026claude}. This was done under the close supervision of the author who also dictated the strategy. We found that perturbative QFT calculations are an ideal use case for agentic AI, since many of the intermediate steps can be offloaded to existing programs used for perturbative calculations, thereby ensuring the correctness of many intermediate steps. Furthermore, we can perform various checks of our calculations at intermediate steps and compare final results against previous results in the literature, leaving us with a high degree of confidence that the results presented here are correct. We discuss various checks of our calculation throughout the paper and checks of our final results in more detail in Section \ref{sec:checks}. While learning how to do this calculation Claude Fable 5.1 was also useful for discussion points about which the author was confused. Further details on our agentic AI strategy can be found in Ref. \cite{LundstrumAgenticAI}.

The structure of this paper is as follows. In Section \ref{sec:theoretical background}, we discuss the four-quark operator bases which we use in this paper. Following this, we discuss the RI renormalization condition and the renormalization of bi-local operators necessary for a lattice QCD calculation of $\epsilon_K$. Section \ref{sec:methods} describes the method, adapted from the one used in Ref.\cite{Gorbahn:2024qpe}, which we use to evaluate the relevant portions of the two-loop amplitudes that contribute to the RI-$\overline{\text{MS}}$ matching factors. In Section \ref{sec:results} we present the results of our calculation. We begin with one-loop calculations entering the matching factors followed by the two-loop bi-local operator calculation. Finally, we give our conclusions in Section \ref{sec:conclusion}.

\section{Theoretical background}
\label{sec:theoretical background}
\subsection{$\Delta S = 1$ operator basis}
\label{sec:deltaS1}
In this subsection we discuss the $\Delta S = 1$ operators that we use for our calculation. We begin with the form of the $\Delta S = 1$ effective weak Hamiltonian
\begin{equation}
H_W^{\Delta S=1}
=
\frac{G_F}{\sqrt{2}}
\left(
\sum_{\substack{q,q'=u,c}}
V_{q's}^*V_{qd}
\sum_{i=1,2} C_i Q_i^{q'q}
-
V_{ts}^*V_{td}
\sum_{i=3}^{6} C_i Q_i
\right).
\label{eq:Heff-four-flav}
\end{equation}
$G_F$ is the Fermi constant and the quantities $V_{ij}$ are Cabibbo–Kobayashi–Maskawa (CKM) matrix elements. The quantities $C_i$ are the Wilson coefficients which encode the short-distance physics, and $Q_i$ are the four-quark low-energy realizations of the weak interactions.

We next define the four-quark operators $Q_i$ that enter $H^{\Delta S = 1}_{\text{eff}}$. First, we have the current-current operators
\begin{equation}
\begin{aligned}
    & Q^{q' \overline{q}}_1 = (\bar{s}_a q'_b)_{V-A} (\bar{q}_b d_a)_{V-A} \\
    & Q^{q' \overline{q}}_2 = (\bar{s}_a q'_a)_{V-A} (\bar{q}_b d_b)_{V-A} \\
\end{aligned}
\end{equation}
Here, $q', q = \{u,c\}$. The QCD penguin operators are defined by
\begin{equation}
\begin{aligned}
Q_3 &= (\bar{s}_a d_a)_{V-A} \sum_{q=u,d,s,c}(\bar{q}_b q_b)_{V-A}, \\
Q_4 &= (\bar{s}_a d_b)_{V-A} \sum_{q=u,d,s,c}(\bar{q}_b q_a)_{V-A}, \\
Q_5 &= (\bar{s}_a d_a)_{V-A} \sum_{q=u,d,s,c}(\bar{q}_b q_b)_{V+A}, \\
Q_6 &= (\bar{s}_a d_b)_{V-A} \sum_{q=u,d,s,c}(\bar{q}_b q_a)_{V+A},
\end{aligned}
\label{eq:Q-QCD-peng}
\end{equation}
In the above, $(\overline{q}q)_{V \pm A} = \overline{q} \gamma_\mu (1 \pm \gamma_5) q$ and $a,b$ are colour indices. Repeated colour indices imply summation. Operators with the colour structure of $Q_1^u$ are called colour mixed and those with the colour structure of $Q_2^u$ are called colour diagonal. The operators with spinor structure $(V-A)\otimes(V-A)$ are called left-left (LL) and those with $(V-A)\otimes(V+A)$ are left-right (LR). We don't consider the electroweak operators because they are suppressed by a factor of $\alpha_{\text{EM}}$ and therefore omitted from lattice QCD calculations of the long-distance contribution to $\epsilon_K$.

The set of operators given above is commonly referred to as the traditional basis. The $\Delta S = 1$ operators can also be organized into the chiral basis in which operators are organized by their transformation properties under chiral rotations. For a detailed discussion of the four-flavour chiral basis we direct the reader to Ref.~\cite{Lehner-Lundstrum}.

\subsection{$\Delta S = 2$ operator basis}
\label{sec:deltaS2}
In this section we describe the local, $\Delta S = 2$ operators that enter the calculation of $\epsilon_K$. These operators will provide the counterterm to cancel the divergence arising from the bi-local operator. The relevant operators are
\begin{equation}
\begin{aligned}
    Q_{LL}^{\Delta S = 2} & = (\overline{s}_a d_a)_{V-A} (\overline{s}_b d_b)_{V-A}, \\
    \tilde Q_{LL}^{\Delta S = 2} & = (\overline{s}_a d_b)_{V-A} (\overline{s}_b d_a)_{V-A}. \\
\end{aligned}
\end{equation}
In this equation, the colour and spinor notation is the same as in the previous subsection. These two operators are related by a Fierz transformation and are therefore not independent operators in four dimensions. In fractional dimensions $D = 4 - 2\epsilon$, the Fierz relations are violated at $O(\epsilon)$. We introduce the following evanescent operator, which we call the Fierz evanescent operator
\begin{equation}
    \tilde Q_{LL} - Q_{LL}  = E_{F}.
\label{eq:Q_LL_fierz_evanescent}
\end{equation}
The matrix elements of the Fierz evanescent operator vanish at tree level after the dimensional regulator $\epsilon$ has been removed. 

\subsection{Regularization-independent renormalization}
In this section we discuss the RI scheme and the factors required to convert between the RI and $\overline{\text{MS}}$ schemes. In this paper we concern ourselves with the symmetric momentum subtraction schemes (SMOM) in which the momentum carried by the external quarks is non-exceptional in the sense that no partial sum of the incoming momenta vanishes ~\cite{Aoki:2007xm,Sturm:2009kb}. Such schemes exhibit better convergence properties than those defined at exceptional momentum points ~\cite{RBC-UKQCD:2008mhs,RBC:2010qam}. 

The RI renormalization condition is imposed by requiring that suitable Green's functions, computed between external off-shell quark and gluon states, in a fixed gauge, coincide with their tree level value \cite{Martinelli:1994ty}. This can also be imposed non-perturbatively and is therefore suitable for lattice QCD calculations. To impose the RI condition we define a projector onto external spin, colour and flavour indices which is applied to the renormalized, amputated Green's function. In this paper we use the following projector
\begin{equation}
P^{ijkl}_{(\gamma_\mu),\alpha\beta\gamma\delta}
=
\frac{
(\gamma^\nu)_{\beta\alpha}(\gamma_\nu)_{\delta\gamma}
+
(\gamma^\nu\gamma^5)_{\beta\alpha}
(\gamma_\nu\gamma^5)_{\delta\gamma}
}{
256\,N_c(N_c+1)
}
\delta_{ab}\delta_{fe}.
\label{eq:RI-gamma-projector}
\end{equation}
This corresponds to the RI/SMOM$(\gamma_\mu, y)$ scheme where $y = \gamma_\mu, \slashed{q}$ refers to the scheme used for the quark wavefunction renormalization factor conversion. We choose this projection scheme for consistency with the scheme used in RBC/UKQCD calculations of the long-distance contribution to $\epsilon_K$ \cite{Bai:2023lkr, Huo:2025bhq}. Here, $N_c = 3$ is the number of colours. Latin indices correspond to colour and the Greek to spinor indices. 

We next discuss the RI $\rightarrow \MSbar$ conversion factors for the quark wavefunction renormalization. In the RI/SMOM scheme these are known to two loops~\cite{Sturm:2009kb, Gorbahn:2010bf, Almeida:2010ns, Gracey:2011fb}. In this paper we work to $O(\alpha_s)$ and so use the one-loop results. We write the conversion factor
\begin{equation}
    C^y_q = 1 + \frac{\alpha_s}{4\pi} \Delta^y_q,
\end{equation}
with
\begin{equation}
\Delta_{q}^{\mathrm{RI'}/\mathrm{MOM}} = \Delta_{q}^{\mathrm{RI}/\mathrm{SMOM}} = -\frac{\xi}{2} \left( N_c - \frac{1}{N_c} \right) +  C_F\,\xi\,\ln\!\big(-p^2/\mu^2_{\overline{\rm MS}}\big),
\label{eq:quark-wf-renorm-SMOM}
\end{equation}
for the RI$'$/MOM and RI/SMOM schemes and
\begin{equation}
\Delta_{q}^{\mathrm{RI/SMOM}_{\gamma_\mu}} = -\frac{1}{2} \left( N_c - \frac{1}{N_c} \right) \left( -1 + \frac{\xi}{2}(3 - C_0) \right) + C_F\,\xi\,\ln\!\big(-p^2/\mu^2_{\overline{\rm MS}}\big),
\label{eq:quark-wf-renorm-SMOM-gamma}
\end{equation}
for the RI/SMO$\text{M}_{\gamma_\mu}$ scheme. Here $p^2 = -\muRI^2$ is the virtuality of the external quarks.

We next discuss how to determine the RI-$\MSbar$ matching factors for the amputated Green's function of a local operator, as we shall do for the $\Delta S = 2$ local operators described in the previous subsection. The following subsection will discuss the RI-$\MSbar$ matching for the bi-local operator. For a more complete discussion, we direct readers to Ref.~\cite{Lehner:2011fz}. Consider the amputated Green's function of an operator renormalized in the $\MSbar$ scheme $O^{\MSbar}$, $\Gamma^{\MSbar} (O^{\MSbar})$. We have
\begin{equation}
    \Gamma^{\MSbar} (O^{\MSbar}) = \Gamma^{\text{tree}} (O) + \sum_X \Upsilon^X \Gamma^{\text{tree}} (X)
\label{eq:MS-greensfunction}
\end{equation}
Here $\Gamma^{\text{tree}}(O)$ is the amputated Green's function of the bare operator $O$ at tree level and $\Gamma^{\text{tree}}(X)$ is the tree level matrix element of the other spinor structures which appear as a result of radiative corrections. The conversion between schemes reads
\begin{equation}
    (C^y_q)^2 \Gamma^{\MSbar} (O^{\MSbar}) = \Gamma^y (O^{\text{RI}}) + \frac{\alpha_s}{4\pi} \Delta r^{\text{RI}\rightarrow \MSbar} \Gamma^y (O^{\text{RI}})
\label{eq:RI-MS-conversion}
\end{equation}
where $y$ denotes the quark wavefunction renormalization scheme. We apply the RI projector to the above and utilize the RI condition $P\big[\Gamma^y (O^{\text{RI}})\big] = P\big[\Gamma^{\text{tree}}(O)\big]$ to find
\begin{equation}
    \Big(1 + 2\frac{\alpha_s}{4\pi}  \Delta^y_q \Big) P\Big[ \Gamma^{\MSbar} (O^{\MSbar})  \Big] = \Big(1 + \frac{\alpha_s}{4\pi} \Delta r^{\text{RI} \rightarrow \MSbar}\Big) P\Big[ \Gamma^{\text{tree}} (O)\Big].
\label{eq:determine-Delta-r}
\end{equation}
We then define the conversion factor $\Delta a^{\text{RI} \rightarrow \MSbar} = \Delta r^{\text{RI} \rightarrow \MSbar} - 2 \Delta^y_q$, which is the projected $\MSbar$ amputated Green's function in units of its tree-level value and is independent of the quark wavefunction renormalization scheme.

\subsection{Bi-local operators in the long-distance contribution to $\epsilon_K$}
\label{sec:renorm-bi-local-description}

Next, we identify the operator combinations that form the long-distance contribution to $\epsilon_k$. We first examine the form of the dispersive part of the $K-\bar K$ mixing matrix which determines $\epsilon_K$
\begin{equation}
M_{\overline{0}0}
=
\langle \overline{K}^0 |
\frac{1}{2}T\left\{
\int d^4x\,H_W^{\Delta S=1}(x,\mu)H_W^{\Delta S=1}(0,\mu)
\right\}_{\mu}
+
H_W^{\Delta S=2}(0,\mu)
|K^0\rangle.
\label{eq:dispersive-K-mix-amp}
\end{equation}
Here, $H_W^{\Delta S=1}(x,\mu)$ is the Hamiltonian in Eq.~\eqref{eq:Heff-four-flav} and $H_W^{\Delta S=2}$ is the local $\Delta S = 2$ Hamiltonian, proportional to $Q_{LL}$, which contains the short-distance contributions to $K -\bar K$ mixing together with the counterterm that removes the divergence of the bi-local term that occurs when $x \rightarrow 0$. To take advantage of CKM matrix unitarity we introduce the quantities
\begin{equation}
    \lambda_i = V_{id} V_{is}^*.
\end{equation}
which satisfy
\begin{equation}
    \lambda_u + \lambda_c + \lambda_t = 0.
\label{eq:CKM-unitarity}
\end{equation}
Next, we examine the product of two $H_W^{\Delta S=1}$ operators in Eq.~\eqref{eq:dispersive-K-mix-amp}. We need to pick out the pieces which contain $\lambda_t$ because this term contributes the CP-violating phase. The usual choice is to use the CKM matrix unitarity to eliminate $\lambda_u = - \lambda_c - \lambda_t$. However, as described in Ref.~\cite{Bai:2023lkr}, it is more convenient for the purposes of a lattice QCD calculation to eliminate $\lambda_c$ and so that is the strategy we use here. Terms proportional to $\lambda_t^2$ are either completely described by short-distance effects, or are numerically irrelevant. This leaves terms proportional to $\lambda_u \lambda_t$ as the target of our calculation. To describe the necessary calculation, we consider the effective $\Delta S = 2$ operator
\begin{equation}
\mathcal{H}^{\Delta S=2}_{W,ut}
=
\frac{G_F^2}{2}\,
\lambda_u \lambda_t
\left\{
\sum_{i=1,2} \sum_{j=1}^6
\int d^4x\,
C_i^{\overline{\mathrm{MS}}}
C_j^{\overline{\mathrm{MS}}}
\Big[ \Big[
\widetilde{Q}_i^{\overline{\mathrm{MS}}}(x)\,
\widetilde{Q}_j^{\overline{\mathrm{MS}}}(0)
\Big]\Big]^{\overline{\mathrm{MS}}}
+
C_{\Delta S = 2}^{\overline{\mathrm{MS}}}
Q_{LL}^{\overline{\mathrm{MS}}}(0)
\right\}.
\label{eq:H-eff-delta-S-2}
\end{equation}
The first term in this equation is the bi-local operator and the second term is the local $\Delta S = 2$ operator which is responsible for the short-distance contribution to $\epsilon_K$. The coefficients $C_i^{\overline{\text{MS}}}$ are the $\Delta S = 1$ Wilson coefficients evaluated in the $\overline{\text{MS}}$ scheme which can be found in Ref.~\cite{Buchalla:1995vs} and $C_{\Delta S = 2}^{\overline{\mathrm{MS}}}$ is the Wilson coefficient of the $\Delta S = 2$ operator. The double brackets surrounding the bi-local operator denote that this is a UV-finite quantity whose divergence has been removed by adding a counterterm proportional to $Q_{LL}$, according to the $\overline{\mathrm{MS}}$ scheme. The operator combinations which enter the bi-local operator are as follows
\begin{align}
  \Big[\Big[\widetilde{Q}_i(x)\widetilde{Q}_j(0)\Big]\Big]
   &= \frac{1}{2}\,T\Big\{
      Q_i^{c\bar c}(x)\big(Q_j^{c\bar c}(0)-Q_j^{u\bar u}(0)\big)
    + \big(Q_i^{c\bar c}(x)-Q_i^{u\bar u}(x)\big)Q_j^{c\bar c}(0)
    \nonumber\\
   &\qquad\quad
    - Q_i^{u\bar c}(x)\,Q_j^{c\bar u}(0) - Q_i^{c\bar u}(x)\,Q_j^{u\bar c}(0)
    \Big\}, \quad (i,j=1,2),
   \label{eq:bilocalCC}\\
   &= \frac{1}{2}\,T\Big\{
      \big[Q_i^{c\bar c}(x)-Q_i^{u\bar u}(x)\big]Q_j(0)
    + Q_j(x)\big[Q_i^{c\bar c}(0)-Q_i^{u\bar u}(0)\big]
    \Big\}, \quad (i=1,2;\; j=3,\ldots,6).
   \label{eq:bilocalCP}
\end{align}

In order to combine the Wilson coefficients evaluated in the $\overline{\text{MS}}$ with the lattice-regularized matrix elements of the four-quark operators, we need to replace the $\overline{\text{MS}}$-renormalized bi-local operator with an RI-renormalized quantity. To do so we impose the RI condition that the sum of the bi-local operator and the $\Delta S = 2$ counterterm required to renormalize its divergence vanishes after applying the RI projectors when evaluated in a Landau gauge-fixed Green's function with off-shell external quarks. Since the $\MSbar$ and RI bi-local operators differ by a term proportional to $Q_{LL}$ we write
\begin{equation}
\int d^4x\,
\left[
\widetilde{Q}_i^{\overline{\mathrm{MS}}}(x)\,
\widetilde{Q}_j^{\overline{\mathrm{MS}}}(0)
\right]^{\overline{\mathrm{MS}}}
=
\int d^4x\,
\left[
\widetilde{Q}_i^{\overline{\mathrm{MS}}}(x)\,
\widetilde{Q}_j^{\overline{\mathrm{MS}}}(0)
\right]^{\mathrm{RI}}
+
Y_{ij}^{\overline{\mathrm{MS}}}
(\mu_{\overline{\mathrm{MS}}},\mu_{\mathrm{RI}})
\,Q_{LL}^{\overline{\mathrm{MS}}}.
\label{eq:MS-biolocal-to-RI}
\end{equation}
This equation determines $Y_{ij}^{\overline{\mathrm{MS}}}$ because the first term on the right-hand side will vanish when the RI condition is imposed at the SMOM kinematic point. Making the projection we find
\begin{equation}
P^{RI} \left[ \int d^4x\,
\left[
\widetilde{Q}_i^{\overline{\mathrm{MS}}}(x)\,
\widetilde{Q}_j^{\overline{\mathrm{MS}}}(0)
\right]^{\overline{\mathrm{MS}}} \right]
=
Y_{ij}^{\overline{\mathrm{MS}}}
(\mu_{\overline{\mathrm{MS}}},\mu_{\mathrm{RI}})
\, P^{RI} \left[ Q_{LL}^{\overline{\mathrm{MS}}} \right].
\end{equation}
We see that on the right side of this equation we have imposed the RI condition on the $O^{\overline{\text{MS}}}_{LL}$. At $O(\alpha_s^0)$ the $\overline{\text{MS}}$ operator corresponds to the RI-renormalized $Q_{LL}$ and no additional factors are required. At $O(\alpha_s)$ we have
\begin{equation}
    Q_{LL}^{\overline{\text{MS}}} = \Big[ 1 + \frac{\alpha_s}{4\pi} \Delta r^{RI \rightarrow \overline{\text{MS}}}_{\Delta S = 2} \Big] Q^{RI}_{LL}.
\end{equation}
Therefore, in order to fully determine the factors $Y_{ij}$ we also need to make a one-loop calculation of the RI/SMOM-$\overline{\text{MS}}$ matching factors for $Q_{LL}$ at our momentum configuration. The actual value of $Y_{ij}^{\overline{\mathrm{MS}}}$ which we report is
\begin{equation}
    Y_{ij}^{\overline{\mathrm{MS}}}
(\mu_{\overline{\mathrm{MS}}},\mu_{\mathrm{RI}}) = P^{RI} \left[ \int d^4x\,
\left[
\widetilde{Q}_i^{\overline{\mathrm{MS}}}(x)\,
\widetilde{Q}_j^{\overline{\mathrm{MS}}}(0)
\right]^{\overline{\mathrm{MS}}} \right] - \frac{\alpha_s}{4\pi} \Delta a^{RI\to\overline{\rm MS}}_{\Delta S=2}\; Y^{(0)}_{ij},
\end{equation}
where $Y^{(0)}$ is the $O(\alpha_s^0)$ contribution to $Y_{ij}^{\overline{\mathrm{MS}}}$. On the right-hand side both terms have the same external quark states, and so the contribution to the RI/SMOM-$\MSbar$ matching factors due to the quark wavefunction renormalization cancels and we therefore just include $ \Delta a^{RI\to\overline{\rm MS}}_{\Delta S=2}$ in the second term. Here, we have made a series expansion of the denominator in $\alpha_s$ and calculate $Y_{ij}^{\overline{\mathrm{MS}}}$ as a subtraction. The difference $O(\alpha_s^2)$ is beyond the perturbative order to which we work and is therefore neglected. 

The two renormalization scales $\muMS$ and $\muRI$ play different roles in this calculation. The quantity $\muMS$ is the usual $\MSbar$ renormalization scale. The second quantity $\muRI$ is determined by the momenta of the external quarks on which the RI condition is imposed. It must be chosen to be far enough below the lattice cut-off scale so that discretization effects are small, but high enough so that the perturbative conversion between the two schemes is applicable. Raising the common value of $\muMS$ and $\muRI$  will result in smaller systematic errors due to truncation of the perturbative series.

The matching factors $Y_{ij}^{\MSbar}$ enter a lattice QCD calculation as follows. First, we use Eq.~\eqref{eq:MS-biolocal-to-RI} to replace the $\MSbar$-renormalized bi-local operator entering the effective Hamiltonian in Eq.~\eqref{eq:H-eff-delta-S-2}. We also replace the RI-renormalized bi-local operator with its lattice counterpart as well as the Wilson coefficients to find
\begin{align}
  \mathcal{H}^{\Delta S=2}_{W,ut} \;=\; \frac{G_F^2}{2}\,\lambda_u\lambda_t
  \Bigg\{ &\sum_{i=1}^{2} \sum_{j=1}^{6}\sum_{x}
    C_i^{\rm Lat} C_j^{\rm Lat}\,
    \Big[\big[\widetilde{Q}^{\rm Lat}_i(x)\,\widetilde{Q}^{\rm Lat}_j(0)\big]\Big]^{\rm RI}
  \nonumber\\
  &+\left( C^{\overline{\rm MS}}_{\Delta S = 2}
     + \sum_{i=1}^{2}\sum_{j=1}^{6} C^{\overline{\rm MS}}_i C^{\overline{\rm MS}}_j\,
       Y^{\overline{\rm MS}}_{ij}(\mu_{\overline{\rm MS}},\mu_{\rm RI}) \right)
     Z^{\rm Lat\to\overline{\rm MS}}_{LL}\, O^{\rm Lat}_{LL}(0) \Bigg\}.
\label{eq:Yij-deliverable}
\end{align}
Since the first term is now evaluated on the lattice the integral over $x$ has been replaced by a discrete sum. In the last term we have also replaced the $\MSbar$ $\Delta S = 2$ operator with its corresponding lattice operator and a scheme conversion coefficient $Z^{\rm Lat\to\overline{\rm MS}}_{LL}$. We see that the matching factors $Y_{ij}^{\MSbar}$ have entered with their indices summed with the corresponding $\Delta S = 1$ Wilson coefficients as a shift in the Wilson coefficient $C_{\Delta S = 2}^{\overline{\mathrm{MS}}}$ corresponding to the $\Delta S = 2$ operator.

\section{Methods}
\label{sec:methods}

The structure of this section is as follows. We first describe the momentum configuration used in the lattice QCD calculation of the long-distance contribution to $\epsilon_K$ \cite{Bai:2023lkr}, which we also use for this perturbative calculation. Next, we list the spinor structures that appear in the two-loop amplitude, along with their corresponding projectors. This is followed by a discussion of the evanescent operators.

\subsection{Momentum configuration}
\label{sec:momentum-configuration}
\begin{figure}[h]
  \centering
  \includegraphics[scale=0.7]{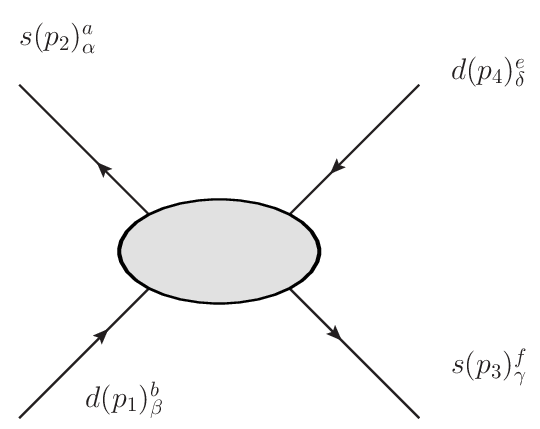} 
   \caption{Momentum and flavour routing for the four-quark amplitude $\Gamma_4$. The quark fields $s$ and $d$ denote the flavour. Greek characters represent spinor indices and Latin characters colour indices.}
    \label{fig:momentum-routing-diag}
\end{figure}
We use the momentum routing in Fig.~\ref{fig:momentum-routing-diag} and choose the conventions used in Eqs. (42) and (43) of Ref.~\cite{Bai:2023lkr}:
\begin{equation}
    p_1 + p_4 = p_2 + p_3,
\end{equation}
\begin{equation}
    p_1^2 = p_2^2 = p_3^2 = p_4^2 = -\mu_{\text{RI}}^2.
\end{equation}
\begin{equation}
  (p_1-p_2)^2 = (p_1-p_3)^2 = (p_2-p_4)^2 = (p_3-p_4)^2 = -\mu_{\rm RI}^2\,,
\end{equation}
\begin{equation}
  (p_1-p_4)^2 = (p_2-p_3)^2 = (p_1+p_4)^2 = -2\mu_{\rm RI}^2\,.
\end{equation}
This choice of momentum is non-exceptional meaning that no partial sum of the momentum vanishes. Such a choice is necessary to control unwanted infra-red effects that can appear when internal quark propagators carry small momentum. Such a choice means that we have three independent external momenta to consider, which we choose as $p_1$, $p_2$ and $p_3$.

\subsection{Tree-level amplitudes}
\label{sec:tree-level-amps}
\begin{figure}[h]
  \centering
  \includegraphics[scale=0.7]{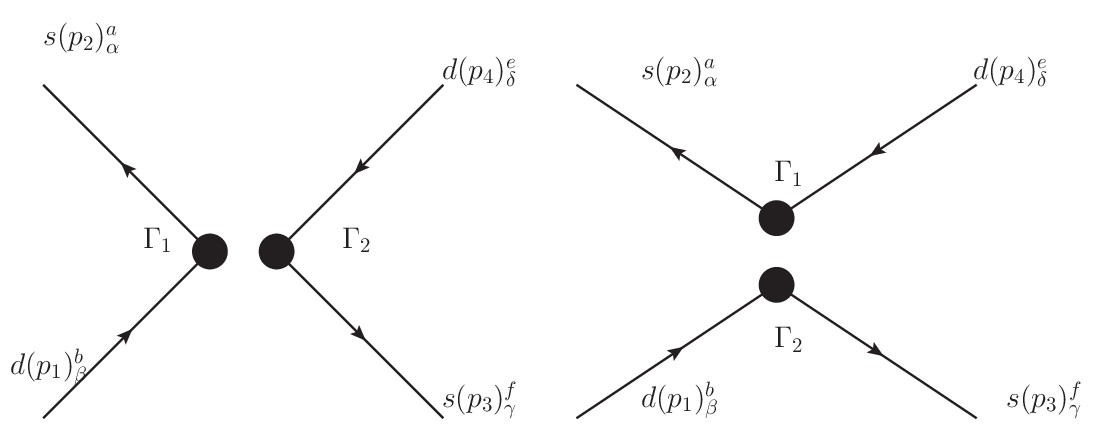} 
   \caption{The two possible tree-level contractions of the four-quark operators with the external quarks. These come in colour-diagonal and colour-mixed combinations}
    \label{fig:momentum-routing-combinations}
\end{figure}
In this section we discuss the tree-level spin-colour structures out of which the finite amplitudes will be constructed. All of the following machinery will apply both to the two-loop calculation of the bi-local operator and the one-loop calculation of the matching factors for the $Q_{LL}$ operator. Because of the identical flavour content in each quark bilinear, there are two possible spinor index contractions which are shown in Fig.~\ref{fig:momentum-routing-combinations}. The corresponding matrix elements of the colour-diagonal $\Delta S = 2$ operators can therefore be written
\begin{equation}
\begin{aligned}
\langle Q \rangle
& =
2\left(
(\gamma^\mu (1 - \gamma_5))_{\alpha\beta}
(\gamma_\mu (1 - \gamma_5))_{\gamma\delta}
\,\delta^{ab}\delta^{fe}
-
(\gamma^\mu (1 - \gamma_5))_{\alpha\delta}
(\gamma_\mu (1 - \gamma_5))_{\gamma\beta}
\,\delta^{ae}\delta^{fb}
\right) \\
& \equiv \langle Q \rangle_2 + \langle Q \rangle_1 
\end{aligned}
\label{eq:Delta-S-2-contraction-colour-diag}
\end{equation}
Here, $\langle Q \rangle_2$ corresponds to the contraction on the left side of Fig.~\ref{fig:momentum-routing-combinations} and $\langle Q \rangle_1$ to the diagram on the right side of Fig.~\ref{fig:momentum-routing-combinations}. We call $\langle Q \rangle_2$ and $\langle Q \rangle_1$ the direct and crossed channels, respectively. The factor of two arises because the final $\Delta S = 2$ structures contain two identical currents. We can repeat this for the colour-mixed $\Delta S = 2$ operator
\begin{equation}
\begin{aligned}
\langle \tilde{Q} \rangle
& =
2\left(
(\gamma^\mu (1 - \gamma_5))_{\alpha\beta}
(\gamma_\mu (1 - \gamma_5))_{\gamma\delta}
\,\delta^{ae}\delta^{fb}
-
(\gamma^\mu (1 - \gamma_5))_{\alpha\delta}
(\gamma_\mu (1 - \gamma_5))_{\gamma\beta}
\,\delta^{ab}\delta^{fe}
\right), \\
& \equiv \langle \tilde{Q} \rangle_2 + \langle \tilde{Q} \rangle_1 
\end{aligned}
\label{eq:Delta-S-2-contraction-colour-mix}
\end{equation}
We next make a note on how the RI/SMOM projectors act in the different Wick contraction channels. Although the spinor indices differ for the Wick contraction channels, those in the RI/SMOM projector in Eq.~\eqref{eq:RI-gamma-projector} against which they are contracted are fixed. Therefore, the action of the RI projector differs for each of the Wick contraction channels. Writing
\begin{equation}
    P^{ijkl}_{(X),\alpha\beta\gamma\delta} \propto \Gamma_{\beta\alpha}\tilde{\Gamma}_{\delta\gamma}\delta_{ab}\delta_{ef},
\end{equation}
We have for the two channels
\begin{equation}
\begin{aligned}
P_{(X)}(\langle Q \rangle_1) &\propto \mathrm{Tr}(\Gamma\gamma^\mu P_L\tilde{\Gamma}\gamma_\mu P_L), \\
P_{(X)}(\langle Q \rangle_2) &\propto \mathrm{Tr}(\Gamma\gamma^\mu P_L)\mathrm{Tr}(\tilde{\Gamma}\gamma_\mu P_L).
\end{aligned}
\label{eq:single-double-RI-traces}
\end{equation}
Explicitly evaluating the spinor parts of traces for the $\gamma_\mu P_L \otimes \gamma^\mu P_L$ structures, we find that the two cases differ by a minus sign, which is compensated for by the relative minus sign between the two terms in Eqs.~\eqref{eq:Delta-S-2-contraction-colour-diag} and \eqref{eq:Delta-S-2-contraction-colour-mix}. 

Because we are only interested in determining the RI/SMOM-$\overline{\text{MS}}$ matching factors, we split the tree-level structures which appear in our amplitudes into two classes: 1) those that contribute to the matching factors and 2) those that vanish under the RI/SMOM projector given in Eq.~\eqref{eq:RI-gamma-projector}. The following tree-level Dirac structures are of class one and contribute to the matching factors
\begin{equation}
Q^s = \gamma^\mu P_L \otimes \gamma_\mu P_L,
\label{eq:Qs-tree-structure}
\end{equation}
\begin{equation}
M_{11}^s = \slashed{p}_1 P_L \otimes \slashed{p}_1 P_L,
\label{eq:M11-tree-structure}
\end{equation}
\begin{equation}
M_{22}^s = \slashed{p}_2 P_L \otimes \slashed{p}_2 P_L,
\end{equation}
\begin{equation}
M_{33}^s = \slashed{p}_3 P_L \otimes \slashed{p}_3 P_L,
\end{equation}
\begin{equation}
M_{12}^s = \frac{1}{2} \left( \slashed{p}_1 P_L \otimes \slashed{p}_2 P_L + \slashed{p}_2 P_L \otimes \slashed{p}_1 P_L \right),
\end{equation}
\begin{equation}
M_{13}^s = \frac{1}{2} \left( \slashed{p}_1 P_L \otimes \slashed{p}_3 P_L + \slashed{p}_3 P_L \otimes \slashed{p}_1 P_L \right),
\end{equation}
\begin{equation}
M_{23}^s = \frac{1}{2} \left( \slashed{p}_2 P_L \otimes \slashed{p}_3 P_L + \slashed{p}_3 P_L \otimes \slashed{p}_2 P_L \right).
\label{eq:M23-tree-structure}
\end{equation}
It should be understood that for these structures and those that follow, the spinor matrix on either side of the tensor product is sandwiched between strange and down quark operators.  For the class two structures we have the following antisymmetric structures:
 \begin{equation}
  A_{12} = \slashed{p}_1 P_L \otimes \slashed{p}_2 P_L - \slashed{p}_2 P_L \otimes \slashed{p}_1 P_L ,
  \label{eq:A12}
  \end{equation}
  \begin{equation}
  A_{13} = \slashed{p}_1 P_L \otimes \slashed{p}_3 P_L - \slashed{p}_3 P_L \otimes \slashed{p}_1 P_L ,
  \label{eq:A13}
  \end{equation}
  \begin{equation}
  A_{23} = \slashed{p}_2 P_L \otimes \slashed{p}_3 P_L - \slashed{p}_3 P_L \otimes \slashed{p}_2 P_L ,
  \label{eq:A23}
  \end{equation}
and
 \begin{equation}
    F^{A}_{12} \;=\; \frac{1}{2}\Big(
       \slashed{p}_{1}\gamma^{\mu}\gamma^{\nu}P_{L}
       \otimes \slashed{p}_{2}\gamma_{\mu}\gamma_{\nu}P_{L}
     \;-\;
       \slashed{p}_{2}\gamma^{\mu}\gamma^{\nu}P_{L}
       \otimes \slashed{p}_{1}\gamma_{\mu}\gamma_{\nu}P_{L}
    \Big),
  \label{eq:FA12}
\end{equation}
  \begin{equation}
    F^{A}_{13} \;=\; \frac{1}{2}\Big(
       \slashed{p}_{1}\gamma^{\mu}\gamma^{\nu}P_{L}
       \otimes \slashed{p}_{3}\gamma_{\mu}\gamma_{\nu}P_{L}
     \;-\;
       \slashed{p}_{3}\gamma^{\mu}\gamma^{\nu}P_{L}
       \otimes \slashed{p}_{1}\gamma_{\mu}\gamma_{\nu}P_{L}
    \Big),
  \label{eq:FA13}
  \end{equation}
  \begin{equation}
    F^{A}_{23} \;=\; \frac{1}{2}\Big(
       \slashed{p}_{2}\gamma^{\mu}\gamma^{\nu}P_{L}
       \otimes \slashed{p}_{3}\gamma_{\mu}\gamma_{\nu}P_{L}
     \;-\;
       \slashed{p}_{3}\gamma^{\mu}\gamma^{\nu}P_{L}
       \otimes \slashed{p}_{2}\gamma_{\mu}\gamma_{\nu}P_{L}
    \Big).
  \label{eq:FA23}
  \end{equation}
These terms include strings of three Dirac matrices which, when $D = 4$, reduce to the antisymmetric analogues of $M_{ij}^s$, as well as structures containing antisymmetric tensors. Terms antisymmetric in the external momenta as well as containing antisymmetric tensors do not contribute to the RI/SMOM-$\overline{\text{MS}}$ matching factors in the scheme considered here.

We next define the projectors onto each of these spinor structures. The projectors corresponding to the class one structures are 
\begin{equation}
\begin{aligned}
& \Pi_\mu(\Gamma_1 P_L \otimes \Gamma_2 P_L) = \frac{1}{2} \operatorname{tr} \big( \gamma^\mu \Gamma_1 \gamma_\mu \Gamma_2 \big), \\
& \Pi_{ij}(\Gamma_1 P_L \otimes \Gamma_2 P_L) =  \frac{1}{2} \operatorname{tr}\big( \slashed{p}_i \Gamma_1 \slashed{p}_j \Gamma_2 \big), 
\end{aligned}
\label{eq:tree-projectors-diag}
\end{equation}
for $i = j = 1,2,3$ and 
\begin{equation}
  \Pi^{S}_{ij}\big(\Gamma_{1}P_{L}\otimes\Gamma_{2}P_{L}\big)
  = \frac{1}{4}\Big[
     \mathrm{tr}\big(\slashed{p}_{i}\,\Gamma_{1}\,\slashed{p}_{j}\,\Gamma_{2}\big)
   + \mathrm{tr}\big(\slashed{p}_{j}\,\Gamma_{1}\,\slashed{p}_{i}\,\Gamma_{2}\big)
     \Big],
\label{eq:tree-projectors-sym}
\end{equation}
with $(ij) = \{(12), (13), (23)\}$. When evaluating these traces, any left-handed projector should be anti-commuted past all gamma matrices to the right side of the matrix chain. Note that the chiral projectors do not appear on the right side of these equations which allows us to avoid $\gamma_5$ trace ambiguities in NDR. The class two structures vanish under the symmetric projection above. We don't provide the projectors required to evaluate the coefficients of the class two structures because these don't contribute in the RI scheme we use.

\subsection{Evanescent operators}
\label{sec:evanescenet-ops}
In this subsection we introduce the evanescent operators: four-quark operators containing three or five Dirac matrices in addition to the left chirality projectors. These arise due to the subtleties associated with defining the Dirac algebra in fractional dimensions.

Dimensional regularization introduces fractional dimensions, in which the definition of $\gamma_5$ and the antisymmetric tensor are ambiguous. For consistency with Refs.~\cite{Buchalla:1995vs, Brod:2010mj, Gorbahn:2024qpe} we choose to work with a naive anti-commuting definition of $\gamma_5$, called naive dimensional regularization (NDR). A consequence of this is that, in addition to the operators discussed in Sections~\ref{sec:deltaS2} and ~\ref{sec:deltaS1}, there are four-quark operators containing additional Dirac matrices which differ from the two gamma matrix structures at $O(\epsilon^n)$ where $n=1$ is relevant at one loop and $n=1,2$ at two loops. These vanish when the regulator is removed and $D = 4$. The difficulty is that these $O(\epsilon^n)$ terms can collide with $1/\epsilon^n$ poles and thereby give non-vanishing contributions to the finite, renormalized amplitudes. We have freedom in choosing the coefficient of the $O(\epsilon^n)$ pieces of the evanescent operators; different choices correspond to different renormalization schemes~\cite{HERRLICH199539}. For our purpose what matters is to choose an evanescent operator scheme consistent with the one used to calculate the $\Delta S = 1$ Wilson coefficients that enter Eq.~\eqref{eq:Heff-four-flav}. These Wilson coefficients are calculated in Ref.~\cite{Buchalla:1995vs}. We also choose an evanescent operator scheme for the $\Delta S = 2$ operators that is consistent with that given in Ref.~\cite{Brod:2010mj} and we follow Ref.~\cite{Gorbahn:2024qpe} in referring to this as the Brod–Gorbahn (BG) evanescent operators basis. 

We first discuss the evanescent operator basis for the $\Delta S = 2$ operators. Instead of working directly in the BG basis, we use the clever strategy devised in Ref.~\cite{Gorbahn:2024qpe}. This is to choose an extension of the convenient Gorbahn-J\"ager-Kvedarait\.e (GJK) evanescent operator basis that vanishes under the projectors defined in the preceding subsection. For consistency with the BG basis, we will eventually make a conversion between the two schemes in the final amputated Green's functions. The GJK evanescent operator basis extended to our case of three independent external momenta is as follows. The first two evanescent operators contain no factors of external momenta.
\begin{equation}
  E_1 = \gamma^\mu\gamma^\nu\gamma^\rho P_L \otimes
           \gamma_\mu\gamma_\nu\gamma_\rho P_L
           + \big((D-10)D+8\big)Q^s ,
\label{eq:O_LL-evanescent-operators_E1}
\end{equation}
\begin{equation}
  E_3 = \gamma^\mu\gamma^\nu\gamma^\rho\gamma^\sigma\gamma^\tau P_L
           \otimes \gamma_\mu\gamma_\nu\gamma_\rho\gamma_\sigma\gamma_\tau P_L
           - (D-2)\big(D((D-26)D+152)-128\big)Q^s .
\label{eq:O_LL-evanescent-operators_E3}
\end{equation}
When the quark fields are included $E_1$ and $E_3$ are colour-diagonal. We call $E_2$ and $E_4$ the analogues colour-mixed structures. The remainder do contain external momenta. We begin with
\begin{align}
  F_{ij} &= \frac{1}{2}\Big(\slashed{p}_i\gamma^\mu\gamma^\nu P_L \otimes
     \slashed{p}_j\gamma_\mu\gamma_\nu P_L + (i\leftrightarrow j)\Big)
     + (D-2)(D-4)\,M^s_{ij} - 4\, p_i\!\cdot\! p_j\, Q^s ,
    \\
  H_{ij} &= \frac{1}{2}\Big(\slashed{p}_i\gamma^\mu\gamma^\nu\gamma^\rho
     \gamma^\sigma P_L \otimes
     \slashed{p}_j\gamma_\mu\gamma_\nu\gamma_\rho\gamma_\sigma P_L
     + (i\leftrightarrow j)\Big) \nonumber\\
   &\quad - \big(D(D-14)+32\big)(D-2)(D-4)\,M^s_{ij}
     + 8(D-8)(D-2)\, p_i\!\cdot\! p_j\, Q^s.
\end{align}
For these two evanescent operators $i,j = \{1,2,3\}$ and they are symmetric in the indices $i,j$. The next two structures are
\begin{align}
  G_1^{(ij)} &= \frac{1}{2}\Big(\slashed{p}_i\slashed{p}_j\gamma^\mu P_L
     \otimes \gamma_\mu P_L
     + \gamma^\mu P_L \otimes \slashed{p}_i\slashed{p}_j\gamma_\mu P_L\Big)
     - p_i\!\cdot\! p_j\, Q^s , \\
  G_3^{(ij)} &= \frac{1}{2}\Big(\slashed{p}_i\slashed{p}_j
     \gamma^\mu\gamma^\nu\gamma^\rho P_L \otimes
     \gamma_\mu\gamma_\nu\gamma_\rho P_L
     + \gamma^\mu\gamma^\nu\gamma^\rho P_L \otimes
     \slashed{p}_i\slashed{p}_j\gamma_\mu\gamma_\nu\gamma_\rho P_L\Big)
     \nonumber\\
   &\quad + \big(D(D-10)+8\big)\, p_i\!\cdot\! p_j\, Q^s .
\end{align}
For these two structures we have $i<j$: (12), (13) and (23). Finally we have
  \begin{align}
  G_2^{(ij)(kl)} &= \frac{1}{2}\Big(
     \slashed{p}_i\slashed{p}_j\gamma^\mu\gamma^\nu\gamma^\rho P_L \otimes
     \slashed{p}_k\slashed{p}_l\gamma_\mu\gamma_\nu\gamma_\rho P_L
     + \big[(ij)\leftrightarrow(kl)\big]\Big) \nonumber\\
   &\quad - (D-4)\big(D(D-14)+32\big)\,\mathcal{M}_{(ij)(kl)}
     - \big(D(D-10)+8\big)\,\mathcal{Q}_{(ij)(kl)}\, Q^s , \\
  G_4^{(ij)(kl)} &= \frac{1}{2}\Big(
     \slashed{p}_i\slashed{p}_j\gamma^\mu P_L \otimes
     \slashed{p}_k\slashed{p}_l\gamma_\mu P_L
     + \big[(ij)\leftrightarrow(kl)\big]\Big)
     + (D-4)\,\mathcal{M}_{(ij)(kl)}
     + \mathcal{Q}_{(ij)(kl)}\, Q^s,
  \end{align}
where
  \begin{align}
  \mathcal{M}_{(ij)(kl)} &\equiv
     (p_j\!\cdot\! p_l)\,M^s_{ik} + (p_i\!\cdot\! p_k)\,M^s_{jl}
   - (p_i\!\cdot\! p_l)\,M^s_{jk} - (p_j\!\cdot\! p_k)\,M^s_{il} , \\
  \mathcal{Q}_{(ij)(kl)} &\equiv
     (p_i\!\cdot\! p_l)(p_j\!\cdot\! p_k)
   - (p_i\!\cdot\! p_j)(p_k\!\cdot\! p_l)
   - (p_i\!\cdot\! p_k)(p_j\!\cdot\! p_l).
  \end{align}
In these structures $i< j$ and $k < l$. 

We also have the evanescent operators associated with the $\Delta S = 1$ operators. The evanescent operators associated with the current-current operators are
\begin{align}
E_1^{q' \overline{q}, {\Delta S = 1}} &=
(\bar{s}_a \gamma_\mu \gamma_\nu \gamma_\rho q'_b)_{V-A}
(\bar{q}_b \gamma^\mu \gamma^\nu \gamma^\rho d_a)_{V-A}
-(16-4\epsilon - 4 \epsilon^2)Q_1 ,
\\
E_2^{q' \overline{q}, {\Delta S = 1}} &=
(\bar{s}_a \gamma_\mu \gamma_\nu \gamma_\rho q'_a)_{V-A}
(\bar{q}_b \gamma^\mu \gamma^\nu \gamma^\rho d_b)_{V-A}
-(16-4\epsilon - 4 \epsilon^2)Q_2.
\label{eq:evanescent_E1_up_basisI}
\end{align}
The evanescent operators associated with the QCD penguin operators are
\begin{align}
E_3^{\Delta S = 1} &=
(\bar{s}_a \gamma_\mu \gamma_\nu \gamma_\rho d_a)_{V-A}
\sum_{q=u,d,s,c}
(\bar{q}_b \gamma^\mu \gamma^\nu \gamma^\rho q_b)_{V-A}
-(16-4\epsilon )Q_3 ,
\\
E_4^{\Delta S = 1} &=
(\bar{s}_a \gamma_\mu \gamma_\nu \gamma_\rho d_b)_{V-A}
\sum_{q=u,d,s,c}
(\bar{q}_b \gamma^\mu \gamma^\nu \gamma^\rho q_a)_{V-A}
-(16-4\epsilon )Q_4 ,
\\
E_5^{\Delta S = 1} &=
(\bar{s}_a \gamma_\mu \gamma_\nu \gamma_\rho d_a)_{V-A}
\sum_{q=u,d,s,c}
(\bar{q}_b \gamma^\mu \gamma^\nu \gamma^\rho q_b)_{V+A}
-(4+4\epsilon)Q_5 ,
\\
E_6^{\Delta S = 1} &=
(\bar{s}_a \gamma_\mu \gamma_\nu \gamma_\rho d_b)_{V-A}
\sum_{q=u,d,s,c}
(\bar{q}_b \gamma^\mu \gamma^\nu \gamma^\rho q_a)_{V+A}
-(4+4\epsilon)Q_6.
\label{eq:evanescent_QCD_penguin}
\end{align}
As stated above, the important thing is to make a choice consistent with that used for the Wilson coefficients determined in Ref.\cite{Buchalla:1995vs}, which are determined with one-loop matching and two-loop running. Because the coefficients of the $O(\epsilon^2)$ pieces only contribute to the finite parts of the amplitudes at two loops, these Wilson coefficients are blind to the choice of $\epsilon^2$ coefficients. Our results do depend on these coefficients, however, and we choose these terms to be consistent with those in Ref. \cite{Brod:2010mj}. For $E_1^{\Delta S = 1}$ and $E_2^{\Delta S = 1}$ this choice is given explicitly in their Eq. (2.10). For the QCD penguin operators, we choose the $\epsilon^2$ terms to vanish for consistency with the operators given in their Eq. (A.27). We check this at vanishing external quark momenta with Eq. (3.44) of Ref. \cite{Brod:2010mj}, after an appropriate change of basis. 

\subsection{Projected amplitudes}
Next, we describe the structure of the two-loop amplitudes in the basis of spinor structures given in the preceding two subsections. To describe the two-loop amplitudes, define the matrix elements of the spinor structures in Eqs.~\eqref{eq:Qs-tree-structure}-~\eqref{eq:M23-tree-structure} as
\begin{equation}
\begin{aligned}
    \langle Q_l \rangle & = \{\langle Q\rangle, \langle M^s_{ij} \rangle\}, \\
    \langle \tilde{Q}_l \rangle & = \{\langle \tilde{Q} \rangle, \langle \tilde{M}^s_{ij} \rangle\}, \\
\end{aligned}
\end{equation}
where again the tilde refers to the mixed colour contraction. All of the spinor structures appear in both of the Wick contraction channels shown in Fig.~\ref{fig:momentum-routing-combinations}. As in Eqs.~\eqref{eq:Delta-S-2-contraction-colour-diag} and \eqref{eq:Delta-S-2-contraction-colour-mix} we distinguish the two channels by writing $\langle Q_l \rangle = \langle Q_l \rangle_2 + \langle Q_l \rangle_1$. We collect these terms into an amplitude for each Wick contraction
\begin{equation}
    \Lambda = \Lambda_1 + \Lambda_2,
\end{equation}
and write for each $\Lambda_a$ with $a = 1,2$
\begin{equation}
\begin{aligned}
\Lambda_a
 & =
\sum_{l=1}^7
\left(
A_{l,a} \langle Q_l \rangle_a
+
\widetilde{A}_{l,a}  \langle \widetilde{Q}_l \rangle_a
\right) + \sum_{k \in (12,13,23)}\Big( A^{A}_{k,a} \langle A_k\rangle_a + \tilde A^A_{k,a}\langle \tilde 
  A_{k}\rangle_a \Big) \\
  & + \sum_{k \in (12,13,23)}\Big( A^{F}_{k,a} \langle F^{A}_{k}\rangle_a + \tilde A^{F}_{k,a}\langle \tilde 
  F^{A}_{k}\rangle_a \Big) \\
 & +\text{linear combinations of evanescent structures}. 
\end{aligned}
\label{eq:Lambda-total-amp}
\end{equation}
The second sum in this equation contains the matrix elements of the antisymmetric spinor structures in Eqs.~\eqref{eq:A12}-\eqref{eq:A23} and the third sum contains the matrix elements of the three-gamma antisymmetric structures in Eqs.~\eqref{eq:FA12}-\eqref{eq:FA23}. The two channels are not independent; the amplitude $\Lambda_1$ is obtained from $\Lambda_2$ by swapping the two $s$ quark legs. This amounts to swapping their spin and colour indices, taking $p_2 \leftrightarrow p_3$ and adding the minus sign associated with the different Wick channel.

To determine the coefficients $A_{l,a}$, we apply the projectors in Eqs.~\eqref{eq:tree-projectors-diag} and \eqref{eq:tree-projectors-sym} to Eq.~\eqref{eq:Lambda-total-amp} to find
\begin{equation}
\begin{aligned}
    \Pi_i(\Lambda_a) & = \sum_{l=1}^7 \Big( A_{l,a} \Pi_i \big(\langle Q_l \rangle_a  \big)  + \tilde A_{l,a} \Pi_i \big(\langle \tilde Q_l \rangle_a  \big) \Big) \\
    & = \sum_{l=1}^7 B_{il} \Big( A_{l,a} \big[ \cdot \big]_{\text{diag}} + \tilde A_{l,a} \big[ \cdot\big]_{\text{mix}} \Big) \\
    & = \Big( C_{i,a} \big[ \cdot \big]_{\text{diag}} + \tilde C_{i,a} \big[ \cdot\big]_{\text{mix}} \Big).
\end{aligned}
\label{eq:B-matrix-def}
\end{equation}
Here, $[\cdot]_{\text{mix/diag}}$ represent the colour index structures appearing in Eqs~\eqref{eq:Delta-S-2-contraction-colour-diag} and \eqref{eq:Delta-S-2-contraction-colour-mix}. In the above the antisymmetric and evanescent structures have vanished under the set of symmetric projectors and we are left with just those relevant spinor structures that contribute to the matching factors. The above makes it clear that after applying the projectors $\Pi_i$ to the diagrams individually, we find the quantities $C_{i,a}$. After inserting one-loop counter terms to cancel sub divergences and two-loop counter terms to cancel the true two-loop divergences the quantities $C_{i,a}$ are finite. We then determine the finite coefficients $A_{l,a} = (B^{-1})_{li} C_{i,a}$. The matrix $B$ is crafted from the spinor structures and is therefore the same in both Wick contraction channels $a=1,2$. The explicit form of the matrix $B$ is given in Table~\ref{tab:Bmatrix}.

\begin{table}[htbp]
      \centering
      \footnotesize
      \renewcommand{\arraystretch}{1.5}
      \setlength{\tabcolsep}{3pt}
      \begin{tabular}{c|ccccccc}
      \hline\hline
       & $\langle Q^s\rangle$ & $\langle M^s_{11}\rangle$ & $\langle M^s_{22}\rangle$ & $\langle M^s_{33}\rangle$ & $\langle M^s_{12}\rangle$ & $\langle M^s_{13}\rangle$ &
      $\langle M^s_{23}\rangle$ \\
      \hline
      $\Pi_\mu$      & $-2D(D-2)$   & $-2(D-2)p^2$ & $-2(D-2)p^2$ & $-2(D-2)p^2$ & $-(D-2)p^2$ & $-(D-2)p^2$ & $0$ \\
      $\Pi_{11}$     & $-2(D-2)p^2$ & $2p^4$  & $-p^4$  & $-p^4$  & $p^4$  & $p^4$  & $p^4$ \\
      $\Pi_{22}$     & $-2(D-2)p^2$ & $-p^4$  & $2p^4$  & $-2p^4$ & $p^4$  & $-p^4$ & $0$ \\
      $\Pi_{33}$     & $-2(D-2)p^2$ & $-p^4$  & $-2p^4$ & $2p^4$  & $-p^4$ & $p^4$  & $0$ \\
      $\Pi^S_{12}$   & $-(D-2)p^2$  & $p^4$   & $p^4$   & $-p^4$  & $2p^4$ & $0$    & $p^4$ \\
      $\Pi^S_{13}$   & $-(D-2)p^2$  & $p^4$   & $-p^4$  & $p^4$   & $0$    & $2p^4$ & $p^4$ \\
      $\Pi^S_{23}$   & $0$          & $p^4$   & $0$     & $0$     & $p^4$  & $p^4$  & $2p^4$ \\
      \hline\hline
      \end{tabular}
      \caption{The matrix $B_{il} = \Pi_i(\langle Q_l\rangle)$ that appears in Eq.~\eqref{eq:B-matrix-def}. Here $p^2 \equiv p_i^2 = -\mu_{\rm RI}^2$. The columns are labelled by
  the spinor structures that contribute to the RI-$\MSbar$ matching factors and the rows by the projectors. Each element corresponds to the result of applying the row projector to
  the column structure. Colour projections are not included. The matrix is symmetric, with $\det B = -64\,(D-2)(D-3)\,(p^2)^{12}$, and is therefore invertible for all $D$ near
  four.}
      \label{tab:Bmatrix}
  \end{table}

\section{Results}
\label{sec:results}
In this section we present our results for the $\overline{\text{MS}}$-renormalized amplitudes arising from the bi-local operator. We begin with the one-loop calculation of the $O(\alpha_s)$ counterterms, finite amplitudes and RI/SMOM-$\overline{\text{MS}}$ matching factors for the $\Delta S = 2$ operator $Q_{LL}$. Following this we give the results for the  one-loop, $O(\alpha_s^0)$ counterterms, finite amplitudes and matching factors for the bi-local operator. Finally, we present the two-loop matching factors for several values of $\muMS$ and $\muRI$. In all of the following we work in four-flavour QCD with massless up, down and strange quarks and a massive charm quark.

To compute these one-loop amplitudes we use the Mathematica packages FeynCalc \cite{Shtabovenko:2023idz, Shtabovenko:2020gxv, Shtabovenko:2016sxi, MERTIG1991345} and LiteRed \cite{Lee:2012cn} to reduce the Feynman loop integrals to a set of Passarino-Veltman scalar master integrals, whose values are known in closed form.

Our two-loop strategy is as follows. The two-loop diagrams are enumerated and classified using QGRAF \cite{qgraf}. The Dirac algebra and projection is carried out using FeynCalc \cite{Shtabovenko:2023idz, Shtabovenko:2020gxv, Shtabovenko:2016sxi, MERTIG1991345}. The projected amplitudes are reduced to a set of master scalar integrals using TARCER \cite{tarcer}, LiteRed \cite{Lee:2012cn} and FIRE6 \cite{Fire6} depending on the integral class. The resulting integrals are evaluated numerically using PySecDec \cite{pysecdec} if they do not have known closed forms, and with TARCER \cite{tarcer} if the closed forms are known. The running of the charm quark mass and strong coupling $\alpha_s$ with the renormalization scale $\mu_{\overline{\text{MS}}}$ is calculated using RunDec \cite{Rundec}. 

Our conventions for the $\MSbar$ renormalization are as follows. For the dimensional regularization we use $D = 4 - 2\epsilon$, working with a naive dimensional regularization (NDR) in which we use an anti-commuting definition of the Dirac matrix $\gamma_5$. For the gluon action we use a covariant gauge fixing term specified by the parameter $\xi$; $\xi = 0$ corresponds to Landau gauge and $\xi = 1$ corresponds to Feynman gauge. We direct readers to Section IIIA of Ref. \cite{Lehner:2011fz} for a more detailed discussion of the subtleties of renormalization in a fixed gauge.

\subsection{One-loop amplitudes: four-quark operators}
\begin{figure}[h]
  \centering
  \includegraphics[scale=0.5]{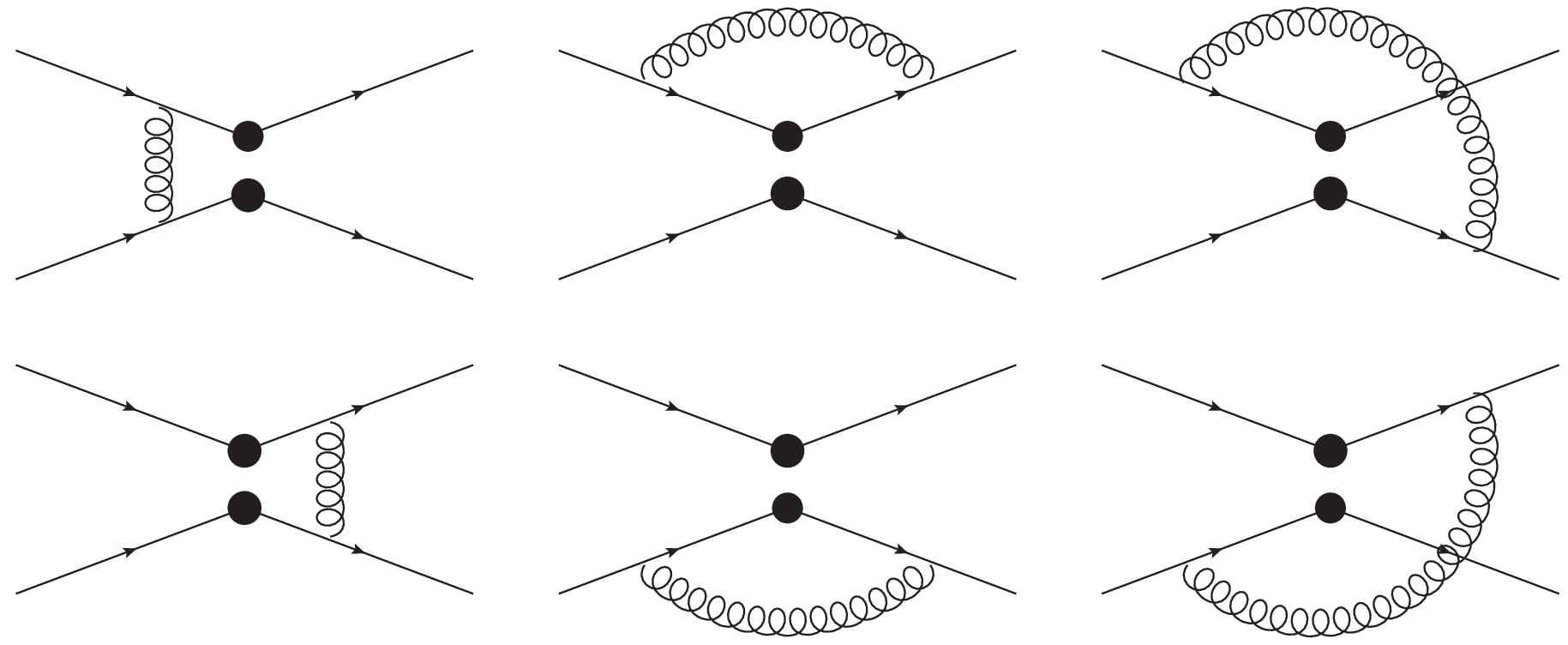} 
   \caption{QCD corrections to the operator $Q_{LL}$ appearing at one loop in QCD perturbation theory. We also consider the same diagrams with the other Wick pairing.}
    \label{fig:current-current-diag}
\end{figure}

We begin by calculating the RI/SMOM-$\overline{\text{MS}}$ matching factors for the $Q_{LL}$ operator that are required for normalizing the $Y_{ij}$ factors, following the methods described in the previous section. The relevant diagrams are shown in Fig. \ref{fig:current-current-diag}.

\subsubsection{Renormalization}

In Section ~\ref{sec:deltaS2}, we listed the following two $\Delta S = 2$ operators
\begin{equation}
\begin{aligned}
    Q_{LL}^{\Delta S = 2} & = (\overline{s}_a d_a)_{V-A} (\overline{s}_b d_b)_{V-A}, \\
    \tilde Q_{LL}^{\Delta S = 2} & = (\overline{s}_a d_b)_{V-A} (\overline{s}_b d_a)_{V-A}.
\end{aligned}
\end{equation}
As was stated in that section, these operators are related by a Fierz transformation and therefore are not independent in four dimensions. Because of the Fierz evanescent operator, it is helpful to consider the radiative corrections to $Q_{LL}$ as mixing two independent operators $Q_{LL}$ and $\tilde Q_{LL}$, then to use the fractional dimension Fierz relation in Eq.~\eqref{eq:Q_LL_fierz_evanescent} to eliminate $\tilde Q_{LL}$ at the end. Calculating the one-loop radiative corrections to $Q_{LL}$, the gluons will mix up the colours and we are left with divergent amplitudes proportional to both the $Q_{LL}$ and $\tilde Q_{LL}$ structures, as well as their corresponding evanescent operators. We define the evanescent operator $E_{2}$ to be the colour-mixed version of the operator $E_1^s$ given in Eq.~\eqref{eq:O_LL-evanescent-operators_E1}. We write the renormalized operator
\begin{equation}
    Q_{LL}^{\overline{\text{MS}}} = 
    \Big[ Z_q^{-2} + Z_{Q}^{\overline{\text{MS}}} \Big] Q_{LL} + b^{\overline{\text{MS}}}_{k} F_{k}.
\end{equation}
Here, $F_k = \{E_1, E_2, E_F\}$, where the Fierz Evanescent operator $E_F$ enters after eliminating the portions of the amplitude proportional to $\tilde Q_{LL}$ using Eq.~\eqref{eq:Q_LL_fierz_evanescent}. A simple power counting reveals that at one loop there are no divergences proportional to spinor structures containing factors of the external momenta. In the above, the quark wavefunction renormalization factor is
\begin{equation}
    Z_q^{2} = 1 - 2C_F \frac{\alpha_s}{4\pi} \frac{\xi}{\epsilon}.
\end{equation}
where $C_F=(N_c^2-1)/(2N_c)$. We find for the remaining renormalization factors
\begin{equation}
    Z_Q^{\overline{\text{MS}}} = \frac{\alpha_s}{4\pi} \frac{1}{\epsilon} \Big[ 3 \Big(1 - \frac{1}{N_c} \Big)\Big],
\end{equation}
\begin{equation}
    b^{\overline{\text{MS}}}_{k} =  \frac{\alpha_s}{4\pi} \frac{1}{\epsilon} \Big( -\frac{1}{2 N_c}, \quad \frac{1}{2}, \quad -3\Big).
\end{equation}
These poles agree with those given in Appendix D of Ref.~\cite{Gorbahn:2024qpe}.

\subsubsection{Finite amplitudes}
Next, we give the renormalized amplitudes in our momentum configuration and the RI/SMOM-${\overline{\text{MS}}}$ matching factors for this operator. To do so, we write the renormalized one-loop amplitude as
\begin{equation}
\begin{aligned}
    \Gamma^{\overline{\text{MS}}} (Q_{LL}^{\overline{\text{MS}}}) = \Gamma^{\text{tree}}(Q_{LL}) & + \frac{\alpha_s}{4\pi} \Big( \Upsilon^\gamma  \Gamma^{\text{tree}}(Q_{LL}) + \sum_{i \leq j=1}^3 \Upsilon^{p}_{ij}  \Gamma^{\text{tree}}(M_{ij}) \Big) \\
    & + \text{antisymmetric terms}.
\end{aligned}
\label{eq:O_LL-one-loop-coeffs}
\end{equation}
The sum in this equation runs over the spinor structures given in Eqs.~\eqref{eq:M11-tree-structure}-\eqref{eq:M23-tree-structure}. For example
\begin{equation}
    M_{11} = (\overline{s}_a \slashed{p}_1 d_a )_{V-A} (\overline{s}_b \slashed{p_1} d_b)_{V-A} / p_1^2.
\end{equation}
We have added here the factor of $p_1^2$ in the denominator to the spinor structures given in Eqs.~\eqref{eq:M11-tree-structure}-\eqref{eq:M23-tree-structure} to keep the dimensionality of the structures the same. We omit terms proportional to the antisymmetric structures in Eqs.~\eqref{eq:A12}-\eqref{eq:FA23} since they do not contribute to the RI/SMOM-$\overline{\text{MS}}$ matching factors, as we shall show below. As noted in Section \ref{sec:tree-level-amps}, the two different Wick channels are not independent. Consider a single structure $S \in \{Q_{LL}, M_{ij}\}$. We have
\begin{equation}
    \Upsilon^S  \Gamma^{\text{tree}}(S) = 2 \Upsilon^S \Big[ S^{\text{(direct)}} \delta^{ab}\delta^{fe} - S^{\text{(crossed)}}_{p_2 \leftrightarrow p_3} \delta^{ae}\delta^{fb} \Big].
\label{eq:O_LL-upsilon-direct-vs-cross-channel}
\end{equation}
Therefore, it is enough for us to specify the coefficients $\Upsilon^S$ in the direct channel, then to utilize Eq.~\eqref{eq:O_LL-upsilon-direct-vs-cross-channel} when computing the matching factors. The coefficients $\Upsilon$ are given in Table~\ref{tab:upsilon-one-loop-OLL}.
\begin{table}[htbp]
    \centering
    \small
    \renewcommand{\arraystretch}{1.6}
    \setlength{\tabcolsep}{4pt}
    \begin{tabular}{c|l}
    \hline\hline
     & $\Upsilon$ \\
    \hline
    $\Upsilon^{\gamma}$ &
    $-9+\dfrac{9}{N_c}-2N_c+\dfrac{C_0}{3}(N_c+1)+6G\Big(1-\dfrac{1}{N_c}\Big)+3\Big(1-\dfrac{1}{N_c}\Big)\ln\!\left(\frac{-p^2}{\mu^2}\right)$ \\
     & $\quad+\,\xi\Big[2N_c-\dfrac{1}{N_c}+2G\Big(1-\dfrac{1}{N_c}\Big)-C_0\Big(\dfrac{2(N_c+1)}{3}-\dfrac{1}{N_c}\Big)-2C_F\,\ln\!\left(\frac{-p^2}{\mu^2}\right)\Big]$ \\
    \hline
    $\Upsilon^{p}_{11}$ & $\dfrac{2(N_c+1)}{3}\Big[(4-C_0)-\xi\,(2-C_0)\Big]$ \\
    \hline
    $\Upsilon^{p}_{22}$ & $\dfrac{2(N_c+1)}{3N_c}\Big[2(2N_c-1)-C_0N_c+\xi\big(C_0(N_c-1)-(2N_c-1)\big)\Big]$ \\
    \hline
    $\Upsilon^{p}_{33}$ & $\dfrac{2(N_c+1)}{3N_c}\Big[2(N_c+1)-C_0N_c+\xi\big(C_0-(N_c+1)\big)\Big]$ \\
    \hline
    $\Upsilon^{p}_{12}$ & $\dfrac{2(N_c+1)}{3N_c}\Big[C_0N_c-2(3N_c-2)+\xi\big((3N_c-2)-2C_0(N_c-1)\big)\Big]$ \\
    \hline
    $\Upsilon^{p}_{13}$ & $\dfrac{2(N_c+1)}{3N_c}\Big[C_0N_c-2(N_c+2)+\xi\big((N_c+2)-2C_0\big)\Big]$ \\
    \hline
    $\Upsilon^{p}_{23}$ & $\dfrac{2(N_c+1)}{3}\Big[2-C_0-\xi\Big]$ \\
    \hline\hline
    \end{tabular}
    \caption{One-loop coefficients entering Eq.~\eqref{eq:O_LL-one-loop-coeffs} at the momentum configuration described in Section ~\ref{sec:momentum-configuration}. In the logarithm
  arguments $p^2 = p_i^2 = -\mu_{\rm RI}^2$. Antisymmetric structures are omitted.}
    \label{tab:upsilon-one-loop-OLL}
  \end{table}
  
Our amplitudes contain the constant $C_0$ defined as
\begin{equation}
    C_0 = (2/3)\Psi'(1/3) - (2\pi/3)^2 \approx 2.34391\,,
\end{equation}
where $\Psi(x)$ is the digamma function. We also encounter Catalan's constant $G$ defined as
\begin{equation}
  G = \sum_{n=0}^{\infty}\frac{(-1)^n}{(2n+1)^2} \approx 0.915966 .
\end{equation}

\subsubsection{Matching factors}
\label{sec:O_LL-one-loop-matching-factors}
To determine the matching factors, we can use the strategy described in Section IV.A of Ref~\cite{Lehner:2011fz}. We consider projectors $P^{\gamma_\mu}$ containing no external momenta. Consider applying this projector to the spinor structures $ \gamma_\mu \Omega_1 \otimes \gamma_\nu \Omega_2$. In this case $\Omega_1 = \Omega_2 = P_L$. Lorentz invariance allows us to write the ansatz
\begin{equation}
    P^{\gamma_\mu} \Big[ \gamma_\mu \Omega_1 \otimes \gamma_\nu \Omega_2 \Big] = g_{\mu \nu} t_1.
\end{equation}
We contract both sides with the metric tensor to determine $t_1$, leading to
\begin{equation}
     P^{\gamma_\mu} \Big[ \gamma_\mu \Omega_1 \otimes \gamma_\nu \Omega_2 \Big] = \frac{1}{4} g_{\mu \nu} P^{\gamma_\mu} \Big[ \gamma_\rho \Omega_1 \otimes \gamma^\rho \Omega_2 \Big].
\label{eq:P_gamma_project}
\end{equation}
To determine the action of the projector onto a spinor structure $M_{ij}$ entering Eq.~\eqref{eq:O_LL-one-loop-coeffs}, we contract both sides of Eq.~\eqref{eq:P_gamma_project} with $p_i^\mu p_j^\nu / p^2$ where $i,j = 1,2,3$. We find
\begin{equation}
     P^{\gamma_\mu} \Big[ \slashed{p}_i \Omega_1 \otimes \slashed{p}_j \Omega_2 / p^2 \Big] = \frac{p_i \cdot p_j}{4 p^2} P^{\gamma_\mu} \Big[ \gamma_\rho \Omega_1 \otimes \gamma^\rho \Omega_2 \Big],
\end{equation}
Implying
\begin{equation}
    P^{\gamma_\mu} \Big[ \Gamma^{\text{tree}}\big(M_{ij}\big) \Big] = \frac{p_i \cdot p_j}{4 p^2}  P^{\gamma_\mu} \Big[ \Gamma^{\text{tree}}(Q_{LL}) \Big]. 
\label{eq:RI-projection-relation-gamma-p-slash}
\end{equation}
In this form it is obvious that the antisymmetric structures in Eqs.~\eqref{eq:A12}-\eqref{eq:FA23} do not contribute to the RI/SMOM-$\overline{\text{MS}}$ matching factors in this scheme. The contributions from Eqs.~\eqref{eq:FA12}-\eqref{eq:FA23} proportional to antisymmetric $\epsilon$ tensors also vanish under the $\gamma_\mu$ projection. This ansatz strategy works for both the single- and double-trace pieces of the RI projection in Eq.~\eqref{eq:single-double-RI-traces}. Applying the $P^{\gamma_\mu}$ to Eq.~\eqref{eq:O_LL-one-loop-coeffs} we find
\begin{equation}
    P^{\gamma_\mu} \Big[ \Gamma^{\overline{\text{MS}}} (Q_{LL}^{\overline{\text{MS}}}) \Big] = P^{\gamma_\mu} \Big[ \Gamma^{\text{tree}}(Q_{LL}) \Big] + \frac{\alpha_s}{4 \pi} \Big( \Upsilon^\gamma + \frac{1}{4} \sum_{i \leq j=1}^3 \Upsilon^{p}_{ij} \frac{p_i \cdot p_j}{p^2}\Big) P^{\gamma_\mu} \Big[ \Gamma^{\text{tree}}(Q_{LL}) \Big]
\label{eq:RI-projected-coefficient-sum}
\end{equation}
Because of our momentum configuration, it happens that the factor $p_i \cdot p_j$ is the same factor after taking $p_2 \leftrightarrow p_3$. Therefore, the projected combination in Eq.~\eqref{eq:RI-projected-coefficient-sum} is also the same in both channels. We can therefore calculate the matching factors with just the $\Upsilon$ coefficients of the direct diagrams. The effect of the different colour routings in each channel is a term $(N_c^2 + N_c)$ multiplying the Upsilon coefficients, which cancels with the same factor in the denominator of the RI projector in Eq.\eqref{eq:RI-gamma-projector}.

Finally, we can use Eq.\eqref{eq:determine-Delta-r} to determine the RI/SMOM-$\overline{\text{MS}}$ matching factors for $Q_{LL}$ as
\begin{equation}
\begin{aligned}
     \Delta a^{RI\to\overline{\rm MS}}_{\Delta S=2} & = \Upsilon^\gamma + \frac14\sum_{i\le j=1}^{3}\Upsilon^p_{ij}\,\frac{p_i\cdot p_j}{p^2}, \\
  \Delta r^{RI\to\overline{\rm MS}}_{\Delta S=2} & = 2\Delta^y_q + \Delta a^{RI\to\overline{\rm MS}}_{\Delta S=2},
\end{aligned}
\end{equation}
where $\Delta^y_q$ is one of the RI$\rightarrow\overline{\text{MS}}$ quark wavefunction conversion factors given in Eqs.~\eqref{eq:quark-wf-renorm-SMOM} or \eqref{eq:quark-wf-renorm-SMOM-gamma}. In Table~\ref{tab:O_LL-Delta-r-factor} we give our results for $\Delta r^{\text{RI}\rightarrow\overline{\text{MS}}}$ in both quark wavefunction RI renormalization schemes.
\begin{table}[htbp]
      \centering
      \small
      \renewcommand{\arraystretch}{1.3}
      \setlength{\tabcolsep}{4pt}
      \begin{tabular}{c|l|l}
      \hline
      $\Delta r^{RI \rightarrow \overline{\text{MS}}}_{\Delta S = 2}/(\alpha_s/4\pi)$ & symbolic & $N_c=3$, $\xi=0$ \\
      \hline
      $(\gamma_\mu,\gamma_\mu)$ &
      $\begin{aligned}
       &-\frac{2(N_c-1)(4-3G)}{N_c} \\
       &-\xi\,\frac{(N_c-1)(1+C_0-4G)}{2N_c} \\
       &+3\Big(1-\frac{1}{N_c}\Big)\ln\!\big(-p^2/\mu^2_{\overline{\rm MS}}\big)
      \end{aligned}$ &
      $\begin{aligned}
       &-1.669471 \\
       &+2.000000\,\ln\!\big(-p^2/\mu^2_{\overline{\rm MS}}\big)
      \end{aligned}$ \\
      \hline
      $(\gamma_\mu,\slashed{q})$ &
      $\begin{aligned}
       &-\frac{(N_c-1)(N_c+9-6G)}{N_c} \\
       &-\xi\,\frac{(N_c-1)\big(C_0(N_c+2)-N_c-4G\big)}{2N_c} \\
       &+3\Big(1-\frac{1}{N_c}\Big)\ln\!\big(-p^2/\mu^2_{\overline{\rm MS}}\big)
      \end{aligned}$ &
      $\begin{aligned}
       &-4.336138 \\
       &+2.000000\,\ln\!\big(-p^2/\mu^2_{\overline{\rm MS}}\big)
      \end{aligned}$ \\
      \hline \hline
      $\Delta a^{RI \rightarrow \overline{\text{MS}}}_{\Delta S = 2}/(\alpha_s/4\pi)$ &
      $\begin{aligned}
       &-\frac{(N_c-1)(N_c+9-6G)}{N_c} \\
       &-\xi\,\frac{(N_c-1)\big(C_0(N_c+2)-3N_c-2-4G\big)}{2N_c} \\
       &+\Big[3\Big(1-\frac{1}{N_c}\Big)-\xi\Big(N_c-\frac{1}{N_c}\Big)\Big] \\
       &\qquad\times\ln\!\big(-p^2/\mu^2_{\overline{\rm MS}}\big)
      \end{aligned}$ &
      $\begin{aligned}
       &-4.336138 \\
       &+2.000000\,\ln\!\big(-p^2/\mu^2_{\overline{\rm MS}}\big)
      \end{aligned}$ \\
      \hline
      \end{tabular}
      \caption{One-loop RI/SMOM$\to\overline{\rm MS}$ matching factors for $Q_{LL}$ at the momentum configuration of Section ~\ref{sec:momentum-configuration}, in units of
    $\alpha_s/4\pi$, with $p^2 = p_i^2 = -\mu_{\rm RI}^2$. The first label denotes the $\gamma_\mu$ projection scheme. The second label the quark wavefunction conversion
    scheme; $\gamma_\mu$ corresponds to Eq.~\eqref{eq:quark-wf-renorm-SMOM-gamma} and $\slashed{q}$ to Eq.~\eqref{eq:quark-wf-renorm-SMOM}. The third row gives the factor $\Delta
  a^{RI \rightarrow \overline{\text{MS}}}_{\Delta S = 2}$ which is the projected one-loop $\overline{\rm MS}$ amplitude of $Q_{LL}$ with no wavefunction conversion applied; this
  is what will actually enter the matching factor for the bi-local operator. In the final column we give the numerical result in Landau gauge where $\xi = 0$.}
    \label{tab:O_LL-Delta-r-factor}
  \end{table}

\subsection{One-loop amplitudes: bi-local operators}
\begin{figure}[t]
  \centering
  \includegraphics[scale=0.7]{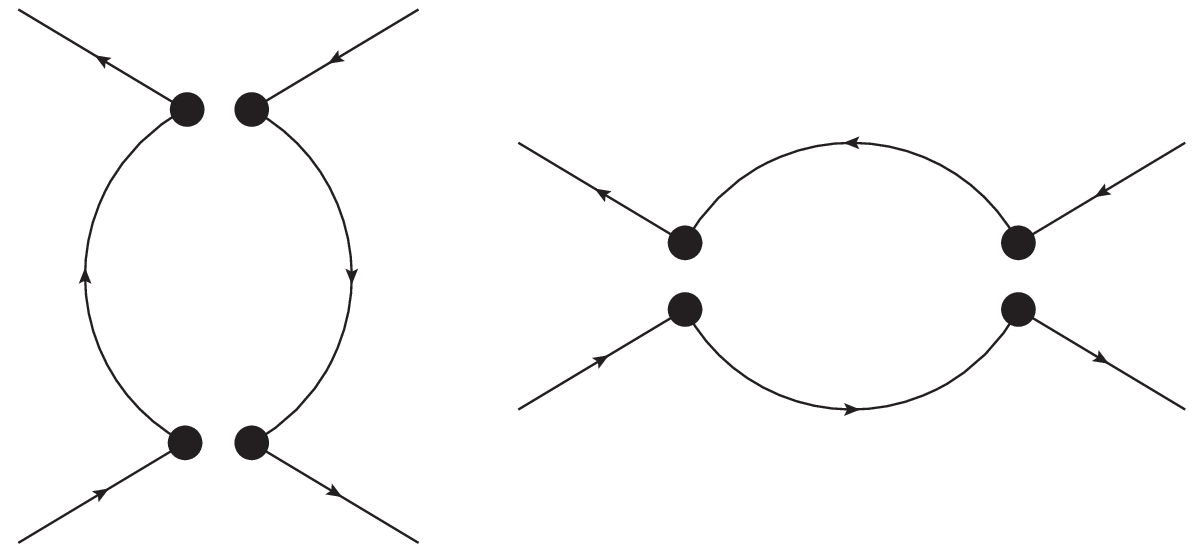} 
   \caption{One-loop diagrams contributing to the bi-local operator. For the case with two current-current insertions the flavour structure is $cc - uc$ and for the case with one current-current operator and one QCD penguin the flavour structure is $cc- uu$.}
    \label{fig:One-loop-diagrams}
\end{figure}

We first consider the contribution to the bi-local operator at one-loop order. The diagrams contributing at this order are shown in Fig.~\ref{fig:One-loop-diagrams}. 

\subsubsection{Renormalization}
\label{sec:bi-local-one-loop-renorm}
Because of the incomplete GIM mechanism, we will encounter divergences proportional to $Q_{LL}$ and evanescent operators when the two $\Delta S = 1$ operators collide. This requires us to introduce counterterms to the Lagrangian in addition to those described in the previous section. We write the $\overline{\text{MS}}$-renormalized bi-local operator as
\begin{equation}
    \Big[ Q_i Q_j \Big]^{\overline{\text{MS}}} = \int d^D x\, T\!\left\{ Q_i^{\overline{\rm MS}}(x)\, Q_j^{\overline{\rm MS}}(0) \right\}
   + Z_{ij}^{\overline{\rm MS}}\, Q_{LL}
   + Z^{\overline{\rm MS}}_{ij,k}\, F_k ,
  \qquad i = 1,2,\; j = 1,\dots,6 .
\label{eq:bi-local-one-loop-counterterms}
\end{equation}
where $Q^{\overline{\text{MS}}}_i$ are the $\Delta S = 1$ operators renormalized in the $\overline{\text{MS}}$ scheme. As in the previous subsection, there are divergences proportional to the evanescent operators $F_k = \{E_1, E_2, E_F\}$. Because the set of $\Delta S = 1$ operators contains colour-mixed and colour-diagonal elements, we will again find amplitudes proportional to $Q_{LL}$ and $\tilde Q_{LL}$ as well as their evanescent partners. We again eliminate $\tilde Q_{LL}$ using the fractional dimension Fierz relation in Eq.~\eqref{eq:Q_LL_fierz_evanescent}, resulting in poles proportional to $E_F$. Power counting reveals that the resulting structures should be dimension eight. Because we consider the difference $cc - uc$ and $cc - uu$, the poles proportional to structures containing factors of the external momenta $q = p_1 - p_2$ and $q = p_1 - p_3$ will cancel in the difference, leaving us only with poles proportional to $m_c^2$. The renormalization factors entering Eq.~\eqref{eq:bi-local-one-loop-counterterms} can be found in Tables~\ref{table:bi-local-one-loop-poles} and \ref{table:bi-local-one-loop-poles-evanescent}. After a change of basis these one-loop poles match those given in Refs.~\cite{Brod:2010mj} and \cite{Herrlich:1996vf}.
\begin{table}[t]
  \centering
  \begin{tabular}{c|cccccc}
  $Z_{ij}\,/\,\dfrac{m_c^2}{4\pi^2\epsilon}$ & $j=1$ & $j=2$ & $j=3$ & $j=4$ & $j=5$ & $j=6$ \\ \hline                                                               
  $i=1$ & $N_c$ & $1$ & $2N_c$ & $2$ & $-2N_c$ & $-2$ \\
  $i=2$ & $1$   & $1$ & $2$    & $2$ & $-2$    & $-2$
  \end{tabular}
  \caption{Coefficients of $Q_{LL}$ in the counterterm of $[Q_iQ_j]$ entering Eq.~\eqref{eq:bi-local-one-loop-counterterms}.}
\label{table:bi-local-one-loop-poles}
\end{table}
\begin{table}[t]
  \centering
  \begin{tabular}{c|ccc}
  $Z_{ij,k}\,/\,\dfrac{m_c^2}{4\pi^2\epsilon}$ & $E_1$ & $E_2$ & $E_F$ \\ \hline
  $(1,1)$ & $0$    & $-N_c/4$ & $N_c$ \\
  $(1,2),(2,1)$ & $0$ & $-1/4$ & $1$ \\
  $(2,2)$ & $-1/4$ & $0$      & $0$ \\
  $(2,4)$ & $0$    & $0$      & $2$ \\
  $(2,6)$ & $0$    & $0$      & $-2$ \\
  \end{tabular}
  \caption{Evanescent counterterms entering Eq.~\eqref{eq:bi-local-one-loop-counterterms}. All pairs not listed have $Z_{ij,k}=0$.}
\label{table:bi-local-one-loop-poles-evanescent}
\end{table}

\subsubsection{Finite amplitudes}
\label{sec:bi-local-one-loop-finite-amps}
Next, we give the one-loop finite amplitudes corresponding to the bi-local operator. These amplitudes are $O(\alpha_s^0)$. We write the matrix elements of the renormalized bi-local operator as
\begin{equation}
\begin{aligned}
  \Gamma^{\overline{\rm MS}}\big([Q_iQ_j]\big)
   & = \frac{m_c^2}{4\pi^2}\Big( \Upsilon^{Q}_{ij}\,\Gamma^{\rm tree}(Q_{LL})
   + \sum_{k\le l=1}^{3} \Upsilon^{p}_{ij,kl}\,\Gamma^{\rm tree}(M_{kl}) \Big) \\
   & + \text{terms which vanish under $P^{\gamma_\mu}$}.
\end{aligned}
\label{eq:bi-local-one-loop-coeffs}
\end{equation}
Because the operators $(Q_1, Q_2)$, $(Q_3, Q_4)$ and $(Q_5, Q_6)$ differ only by colour indices within the pairs, we can write the coefficients in the following form
\begin{equation}
\begin{aligned}
    \Upsilon^{Q}_{ij} & = \tau_{ij} \phi^{Q}_{j} \\
    \Upsilon^{p}_{ij,kl} & = \tau_{ij}\, w_{kl}\, \phi^{p}_{j}
\end{aligned}
\end{equation}
In the above, $\tau_{ij}$ is the colour table and $w_{kl}$ records how the momentum transfers in the two Wick contractions decompose into the $M_{kl}$ basis. In matrix form they are
\begin{equation}
    \tau_{ij} =
  \begin{pmatrix}
  N_c & 1 & N_c & 1 & N_c & 1 \\
  1   & 1 & 1   & 1 & 1   & 1
  \end{pmatrix},
  \qquad
  w_{kl} =
  \begin{pmatrix}
   1 & -1 & -1 \\
  -1 & \tfrac12 & 0 \\
  -1 & 0 & \tfrac12
  \end{pmatrix}.
\end{equation}
In Table \ref{table:bi-local-one-loop} we give the finite, one-loop amplitudes for the bi-local operator. We use the quantities
\begin{equation}
  \bar\beta = \sqrt{1 - \frac{4m_c^2}{p^2}}, \qquad \Lambda = \ln\left(\frac{\bar\beta + 1}{\bar\beta - 1}\right).
\end{equation}

\begin{table}[htbp]
    \centering
    \small
    \renewcommand{\arraystretch}{1.3}
    \setlength{\tabcolsep}{4pt}
    \begin{tabular}{c|l|l}                                                                                                 
    \hline                                                                                                                                                                    
    block & $\phi^{Q}_{j}/(m_c^2/4\pi^2)$ & $\phi^{p}_{j}/(m_c^2/4\pi^2)$ \\ \hline
    $j=1,2$ &
    $\begin{aligned}[t] &\ln\left(\frac{m_c^2}{\mu^2}\right) - \frac13 - \frac{m_c^2}{3p^2}
     - \frac{p^2-4m_c^2}{3m_c^2}\,\bar\beta\Lambda \\ &- \frac{(m_c^2-p^2)^3}{3m_c^2p^4}\,\ln\left(1-\frac{p^2}{m_c^2}\right) \end{aligned}$ &
    $\begin{aligned}[t] &\frac43 + \frac{4m_c^2}{3p^2}
     - \frac{2(2m_c^2+p^2)}{3m_c^2}\,\bar\beta\Lambda \\ &+ \frac{2(m_c^2-p^2)^2(2m_c^2+p^2)}{3m_c^2p^4}\,\ln\left(1-\frac{p^2}{m_c^2}\right) \end{aligned}$ \\[3ex] \hline
    $j=3,4$ &
    $\begin{aligned}[t] &\frac23 + 2\ln\left(\frac{m_c^2}{\mu^2}\right) + \frac{2p^2}{3m_c^2}\ln\left(\frac{-p^2}{m_c^2}\right) \\ &+
    \frac23\left(1-\frac{p^2}{m_c^2}\right)\bar\beta\Lambda \end{aligned}$ &
    $\begin{aligned}[t] &-\frac83 - \frac{2p^2}{3m_c^2}\ln\left(\frac{-p^2}{m_c^2}\right) \\ &+ \frac23\left(2+\frac{p^2}{m_c^2}\right)\bar\beta\Lambda \end{aligned}$
    \\[3ex] \hline
    $j=5,6$ & $2\left(1 - \ln\left(\frac{m_c^2}{\mu^2}\right) - \bar\beta\Lambda\right)$ & $0$ \\
    \hline
    \end{tabular}
    \caption{Finite parts $\phi^{Q}_j$ and $\phi^{p}_j$ of Eq.~\eqref{eq:bi-local-one-loop-coeffs}, in units of $m_c^2/4\pi^2$.}
    \label{table:bi-local-one-loop}
  \end{table}

\subsubsection{Matching factors}
Next we report the one-loop RI/SMOM-$\overline{\text{MS}}$ matching factors. Since the tree-level structures entering Eq.~\eqref{eq:bi-local-one-loop-coeffs} are the same as in the previous section, we can use the same method described in Section ~\ref{sec:O_LL-one-loop-matching-factors} to determine the matching factors for the bi-local operator. Because this calculation is $O(\alpha_s^0)$, there are no quark wavefunction renormalization factors or dependence on the gauge fixing parameter $\xi$. After applying the RI projector $P^{\gamma_\mu}$ and completing the sum over $k,l$ we find
 \begin{equation}
  Y_{ij}^{(0), \overline{\text{MS}}} (\mu_{ \overline{\text{MS}}}, \mu_{\text{RI}}) = \frac{P^{(\gamma^\mu)}\big[\Gamma^{\overline{\rm MS}}([Q_iQ_j])\big]}{P^{(\gamma^\mu)}\big[\Gamma^{\rm tree}(Q_{LL})\big]}\Bigg|_{p^2=-\mu_{\rm RI}^2}
  = \frac{m_c^2}{4\pi^2}\,\tau_{ij}\,\Big(\phi^{Q}_{j} + \tfrac14\,\phi^{p}_{j}\Big)\Big|_{p^2=-\mu_{\rm RI}^2}
  \equiv \frac{m_c^2}{4\pi^2}\,\tau_{ij}\,\chi_j ,
\label{eq:bi-local-RI-projection}
\end{equation}
We give the values of $\chi_j$ in Table~\ref{table:bi-local-one-loop-matching-factors}. In Ref.~\cite{Buchalla:1995vs} Eq.(12.64), they have eliminated $\lambda_u$ rather than $\lambda_c$ as we have. For $j=3,...,6$, the difference is just an overall minus sign. For $j = 1,2$ their flavour combination is thus $uu-uc$ rather than our $cc - uc$. Since at $p^2 = 0$ the $uu$ integral is scaleless, it vanishes in dimensional regularization and therefore we have confirmed that we reproduce the current-current results of Ref.~\cite{Buchalla:1995vs} by considering just our $uc$ piece.
  \begin{table}[htbp]
    \centering
    \small
    \renewcommand{\arraystretch}{1.3}
    \setlength{\tabcolsep}{4pt}
    \begin{tabular}{c|l}
    \hline
    block & $\chi_j = \phi^{Q}_{j} + \tfrac14\,\phi^{p}_{j}$ \\ \hline
    $j=1,2$ &
    $\ln\left(\frac{m_c^2}{\mu^2}\right)
     + \frac{2m_c^2-p^2}{2m_c^2}\,\bar\beta\Lambda
     + \frac{(m_c^2-p^2)^2}{2m_c^2\,p^2}\,\ln\left(1-\frac{p^2}{m_c^2}\right)$ \\[2ex] \hline
    $j=3,4$ &
    $2\ln\left(\frac{m_c^2}{\mu^2}\right)
     + \frac{p^2}{2m_c^2}\,\ln\left(\frac{-p^2}{m_c^2}\right)
     + \frac{2m_c^2-p^2}{2m_c^2}\,\bar\beta\Lambda$ \\[2ex] \hline
    $j=5,6$ &
    $2\left(1-\ln\left(\frac{m_c^2}{\mu^2}\right)-\bar\beta\Lambda\right)$ \\
    \hline
    \end{tabular}
    \caption{The $(\gamma^\mu)$-projected finite amplitudes $\chi_j$. The matching factor for the pair $(Q_i,Q_j)$ is
    given by Eq.~\eqref{eq:bi-local-RI-projection} evaluated at $p^2=-\mu_{\rm RI}^2$.}
  \label{table:bi-local-one-loop-matching-factors}
  \end{table}

\subsection{Two-loop amplitudes}
\begin{figure}[t]
  \centering
  \includegraphics[scale=0.5]{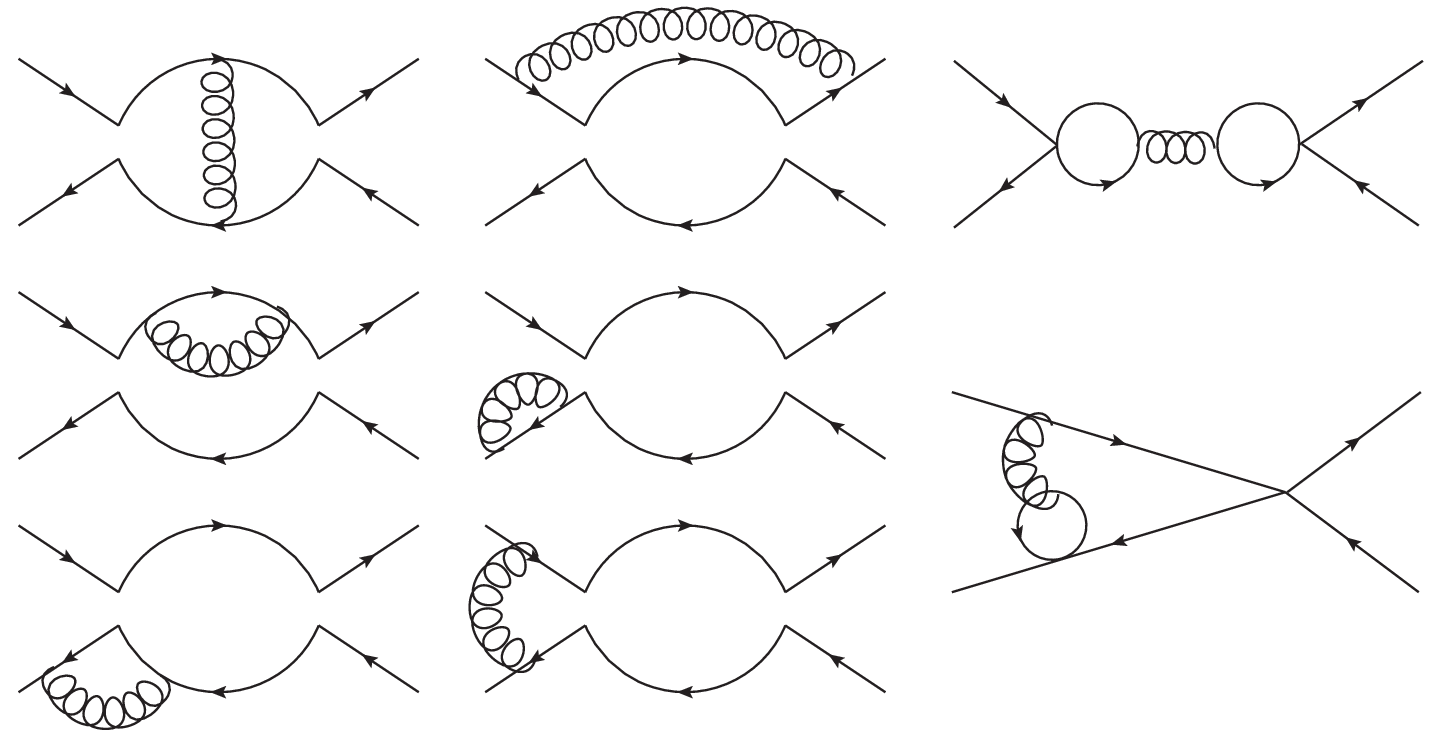} 
   \caption{Classes of two-loop diagrams contributing to the RI-$\MSbar$ matching factors $Y_{ij}$. The diagrams not shown here vanish under RI projection and are therefore not considered.}
    \label{fig:two-loop-diagrams}
\end{figure}

In this section we calculate the two-loop amplitudes to determine the counterterms and finite amplitudes that contribute to the RI/SMOM-$\overline{\text{MS}}$ amplitudes at $O(\alpha_s)$. The classes of diagrams we evaluate are shown in Fig. \ref{fig:two-loop-diagrams}. In addition to these bare two-loop diagrams, we also need to evaluate the one-loop diagrams where one-loop counterterms have been inserted to cancel subdivergences. The counterterms corresponding to the quark loop are taken from those determined in Section \ref{sec:bi-local-one-loop-renorm} and those corresponding to the $\Delta S = 1$ vertices were determined in Ref.~\cite{Lehner-Lundstrum}. 

\subsubsection{Renormalization}
In this section, we expand the definition of the $\overline{\text{MS}}$-renormalized bi-local operator in Eq.~\eqref{eq:bi-local-one-loop-counterterms} to include the two-loop counterterms.
\begin{equation}
    \Big[Q_iQ_j\Big]^{\overline{\rm MS}}
   = \int d^Dx\, T\Big\{Q_i^{\overline{\rm MS}}(x)\,Q_j^{\overline{\rm MS}}(0)\Big\} \,
   + \, \Big[ Z^{(0)}_{ij}  + \frac{\alpha_s}{4\pi}\, Z^{(1)}_{ij} \Big] Q_{LL}^{\overline{\rm MS}}
   \,+\, \Big[Z^{(0)}_{ij,k} + \frac{\alpha_s}{4\pi}\, Z^{(1)}_{ij,k}\Big] F_k^{\overline{\rm MS}},
\label{eq:bi-local-two-loop-counterterms}
\end{equation}
with $i=1,2,\; j=1,\dots,6$. In the above the superscript $0$ in $Z^{(0)}$ denotes the counterterms coming from the one-loop, $O(\alpha_s^0)$ diagrams discussed in Section \ref{sec:bi-local-one-loop-renorm}. The superscript $(1)$ denotes the two-loop counterterms which we discuss in this section. Focusing on $Z^{(1)}$ we have
\begin{equation}
\begin{aligned}
      Z^{(1)}_{ij} & = \frac{\bar m_c^2}{4\pi^2}\Big(\frac{a^{(2)}_{ij}}{\epsilon^2} + \frac{a^{(1)}_{ij}}{\epsilon}\Big), \\
      Z^{(1)}_{ij,k} & = \frac{\bar m_c^2}{4\pi^2}\Big(\frac{a^{(2)}_{ij,k}}{\epsilon^2} + \frac{a^{(1)}_{ij,k}}{\epsilon}\Big).
\end{aligned}
\label{eq:bi-local-two-loop-Z-factors}
\end{equation}
In these equations $\overline{m}_c$ denotes the $\overline{\text{MS}}$-renormalized charm quark mass. We also need as a mass counter term the charm mass renormalization factor
\begin{equation}
    Z_m = 1 - \frac{\alpha_s}{4\pi} \frac{3 C_F}{\epsilon}.
\end{equation}
Recall that our strategy described in Section \ref{sec:methods} uses a set of projectors which annihilate the evanescent operators. Therefore, the terms $Z^{(1)}_{ij,k}$ with $k = E_1, E_2$ are not determined by our calculation. We do determine the poles corresponding to the Fierz evanescent operators which enter through the elimination of the colour-mixed structures proportional to $\tilde Q_{LL}$. The factors $a$ entering Eq.~\eqref{eq:bi-local-two-loop-Z-factors} are given in Table \ref{table:bi-local-two-loop-counterterms}. We have checked that these results reproduce the anomalous dimension matrices given in Eq. (3.18) of Ref. \cite{Brod:2010mj} and Eq. (89b) of \cite{Herrlich:1996vf}, after rotation to the appropriate $Q_{\pm}$ basis.
 \begin{table}[htbp]
    \centering
    \footnotesize
    \renewcommand{\arraystretch}{1.6}
    \setlength{\tabcolsep}{3pt}
    \begin{tabular}{c|cccccc}
    \hline\hline
     & $j=1$ & $j=2$ & $j=3$ & $j=4$ & $j=5$ & $j=6$ \\
    \hline
    $a^{(2)}_{1j}$ & $\frac{3N}{2}-\frac{3N^2}{2}$ & $3-\frac{3}{N}$ & $3N-3N^2$ & $6-\frac{6}{N}$ & $3N^2-3N$ & $6N-6$ \\
    $a^{(2)}_{2j}$ & $3-\frac{3}{N}$ & $\frac{9}{2}-\frac{3N}{2}-\frac{3}{N}$ & $6-\frac{6}{N}$ & $9-3N-\frac{6}{N}$ & $0$ & $6N-6$ \\
    $a^{(1)}_{1j}$ & $\frac{11N^2}{4}+\frac{41N}{4}-13$ & $6-\frac{3N}{2}-\frac{9}{2N}$ & $\frac{11N^2}{2}+\frac{17N}{2}-14$ & $9N-\frac{9}{N}$ &
    $8-\frac{11N^2}{2}-\frac{5N}{2}$ & $4N+6-\frac{10}{N}$ \\
    $a^{(1)}_{2j}$ & $6-\frac{3N}{2}-\frac{9}{2N}$ & $\frac{29N}{4}-\frac{11}{4}-\frac{9}{2N}$ & $18-3N-\frac{15}{N}$ & $\frac{29N}{2}+\frac{1}{2}-\frac{15}{N}$ &
    $3N+7-\frac{10}{N}$ & $12-2N-\frac{10}{N}$ \\
    \hline
    $a^{(2)}_{1j,E_F}$ & $-\frac{3N^2}{2}$ & $-\frac{3}{N}$ & $3N$ & $6$ & $-3N$ & $-6$ \\
    $a^{(2)}_{2j,E_F}$ & $-\frac{3}{N}$ & $\frac{9}{2}$ & $6$ & $-3N-\frac{6}{N}$ & $0$ & $6N$ \\
    $a^{(1)}_{1j,E_F}$ & $\frac{11N^2}{4}-13$ & $-\frac{3N}{2}-\frac{9}{2N}$ & $\frac{17N}{2}$ & $0$ & $-\frac{5N}{2}$ & $6$ \\
    $a^{(1)}_{2j,E_F}$ & $-\frac{3N}{2}-\frac{9}{2N}$ & $-\frac{11}{4}$ & $18$ & $\frac{29N}{2}-\frac{15}{N}$ & $7$ & $-2N-\frac{10}{N}$ \\
    \hline\hline
    \end{tabular}
    \caption{Two-loop counterterm coefficients of Eq.~\eqref{eq:bi-local-two-loop-Z-factors} in units of $\bar m_c^2/(4\pi^2)$ with $\overline{\rm MS}$ charm mass, given after removing the quark wavefunction renormalization factor.  The $E_1$,
  $E_2$ columns are not determined by the projected calculation.}
    \label{table:bi-local-two-loop-counterterms}
  \end{table}

\subsubsection{Conversion to Brod-Gorbahn basis}
\label{sec:convert-BG-evanescent-basis}

In this section we make a change of evanescent scheme so that our final results are consistent with the BG scheme used in Ref.\cite{Brod:2010mj}. As was described in Section \ref{sec:evanescenet-ops}, we choose a set of evanescent operators that vanish under our set of projectors. The change of evanescent scheme amounts to a finite shift in the amplitude proportional to $\overline{m}_c^2$ and independent of logs containing $\mu_{\text{RI}}$ and $\mu_{\overline{\text{MS}}}$. The evanescent operators in the BG basis are
\begin{align}
E_1^{(2),\text{BG}} &= (\bar{s}^i \gamma^\mu\gamma^\nu\gamma^\rho\gamma^\sigma\gamma^\tau  P_L d^j)(\bar{s}^k  \gamma_\mu\gamma_\nu\gamma_\rho\gamma_\sigma\gamma_\tau P_L d^l) - \left( 256 - 224\epsilon - \frac{108\,816}{325}\epsilon^2 \right) Q, \\
E_2^{(2),\text{BG}} &= (\bar{s}^i \gamma^\mu\gamma^\nu\gamma^\rho\gamma^\sigma\gamma^\tau P_L d^l)(\bar{s}^k \gamma_\mu\gamma_\nu\gamma_\rho\gamma_\sigma\gamma_\tau P_L d^j) - \left( 256 - 224\epsilon - \frac{108\,816}{325}\epsilon^2 \right) (Q + E_F),
\end{align}
The difference of these relative to the evanescent operators given previously in Eq.~\eqref{eq:O_LL-evanescent-operators_E3} is at $O(\epsilon^2)$. To change to the BG scheme we use
\begin{align}
E_3 &= E_1^{(2),\mathrm{BG}} + \kappa \epsilon^2 Q,  \\
E_4 &= E_2^{(2),\mathrm{BG}} + \kappa \epsilon^2 (Q + E_F),
\end{align}
with $\kappa = -62016/325$. The finite shifts to the amplitudes between evanescent schemes will occur due to these  $O(\epsilon^2)$ terms colliding with $1/\epsilon^2$ poles proportional to the evanescent operators $E_3$ and $E_4$ in our calculations. The amputated Green's functions differ as 
\begin{equation}
  \big\langle [Q_iQ_j]^{\overline{\rm MS},\rm GJK}\big\rangle
   = \big\langle [Q_iQ_j]^{\overline{\rm MS},\rm BG}\big\rangle
   + \frac{\alpha_s}{4\pi} \frac{\bar m_c^2}{4\pi^2}\,\kappa\Big[c^{(2,2)}_{ij,E_3}\,\langle Q_{LL}\rangle^{\rm tree}
   + c^{(2,2)}_{ij,E_4}\big(\langle Q_{LL}\rangle^{\rm tree}+\langle E_F\rangle^{\rm tree}\big)\Big],
  \label{eq:bg-conv-bilocal}
\end{equation}
Where 
\begin{equation}
  c^{(2,2)}_{ij,E_n} =\frac12\Big[\sum_{m=1,2} Z^{(0)}_{ij,E_m} Z^{(1,1)}_{E_mE_n}
   + \sum_k \big( Z^{(1,1)}_{Q_iE^{\Delta S = 1}_k}\, Z^{(0)}_{[E^{\Delta S = 1}_kQ_j],E_n} + (i\leftrightarrow j)\big)\Big]
  \qquad n=3,4 .
  \label{eq:BG-conversion-eps-square-terms}
\end{equation}

In this equation, the superscript $Z^{(0)}$ denotes the counterterm coefficient of a $1/\epsilon$ pole coming from the $O(\alpha_s^0)$ quark loop, and the superscript $Z^{(1,1)}$ denotes the counterterm coefficient of a $O(\alpha_s)$, $1/\epsilon$ pole. We now discuss each of these factors, beginning with those $O(\alpha_s^0)$. The factor $Z^{(0)}_{ij,E_m}$ is the counterterm coefficient multiplying a three-gamma, $\Delta S = 2$ evanescent operator $E_{1,2}$ which appears after completing the bi-local quark loop with vertices made from the $\Delta S = 1$ operators $Q_i$ and $Q_j$. The matrix $Z^{(0)}_{ij,E_m}$ is given in Table \ref{table:bi-local-one-loop-poles-evanescent}. The next factor $Z^{(0)}_{[E^{\Delta S = 1}_k Q_j],E_n}$ is the counterterm coefficient multiplying the five-gamma, $\Delta S = 2$ evanescent operator $E_{3,4}$ which occurs after completing the remaining quark loop which has the $\Delta S = 1$ evanescent operator $E^{\Delta S = 1}_k$ at one vertex and the $\Delta S = 1$ operator $Q_j$ at the other vertex. The matrix $Z^{(0)}_{[E^{\Delta S = 1}_k Q_j],E_n}$ has the same entries as $Z^{(0)}_{kj,E_m}$ with $E_3 \leftrightarrow E_1$ and $E_4\leftrightarrow E_2$. This is because the loop diagrams for the two cases are identical except for the number of gamma matrices at the vertices, which just changes the number of gamma matrices in the final structure. We have verified that this is the case. The factor $Z^{(1,1)}_{Q_i E^{\Delta S = 1}_k}$ denotes the counterterm coefficient proportional to the three-gamma, $\Delta S = 1$ evanescent operator $E^{\Delta S = 1}_k$ which appears after completing a gluon loop correction to the $\Delta S = 1$ operator $Q_i$. For $Z^{(1,1)}_{Q_i E^{\Delta S = 1}_k}$ we use the results calculated in Ref.~\cite{Lehner-Lundstrum}, which we copy here
\begin{equation}
    Z^{(1,1)}_{Q_iE^{\Delta S = 1}_k} = \begin{pmatrix} \dfrac{N_c}{4}-\dfrac{1}{2N_c} & \dfrac{1}{4}\\[6pt] \dfrac{1}{2} & -\dfrac{1}{2N_c}\end{pmatrix},
  \qquad i,k = 1,2.
\end{equation}
The factor $Z^{(1,1)}_{E_m E_n}$ is the counterterm coefficient which occurs after completing the gluon loop that corrects the $\Delta S = 2$ evanescent operators $E_{1,2}$. This results in the five-gamma evanescent operators $E_{3,4}$. Here too we find that $Z^{(1,1)}_{E_m E_n}$ has the same entries as $Z^{(1,1)}_{Q_i E^{\Delta S = 1}_k}$ with $E_1 \leftrightarrow Q_2$,  $E_2 \leftrightarrow Q_1$,  $E_3 \leftrightarrow E_2^{\Delta S = 1}$ and $E_4 \leftrightarrow E_1^{\Delta S = 1}$ for the same reason as was given above for $Z^{(0)}_{[E^{\Delta S = 1}_k Q_j],E_n}$. We have verified this against the results given in Appendix D of Ref. \cite{Gorbahn:2024qpe}. Finally, the second term is symmetrized in $i,j$ because the $\Delta S = 1$ divergence can appear at either vertex.

The factors $c^{(2,2)}_{ij,E_n} = 0$ for $j = 3,...,6$, because the flow of fermion lines through the diagram are such that we don't form operators with five gamma matrices in each quark bilinear.

\subsubsection{Finite amplitudes}
We next discuss the two-loop finite amplitudes at $O(\alpha_s)$. We use the same form of the finite amplitudes as in Eq.~\eqref{eq:bi-local-one-loop-coeffs} 
\begin{equation}
\begin{aligned}
  \Gamma^{\overline{\rm MS}}\big([Q_iQ_j]\big)
   & = \frac{\bar m_c^2}{4\pi^2}\Big( \Upsilon^{Q}_{ij}\,\Gamma^{\rm tree}(Q_{LL})
   + \sum_{k\le l=1}^{3} \Upsilon^{p}_{ij,kl}\,\Gamma^{\rm tree}(M_{kl}) \Big) \\
   & + \text{terms which vanish under $P^{\gamma_\mu}$},
\end{aligned}
\end{equation}
Here, we use the $\overline{\text{MS}}$-renormalized charm mass $\overline{m}_c =\overline{m}_c(\mu_{\overline{\text{MS}}})$. We write each of the coefficients as  $\Upsilon = \Upsilon^{(0)} + (\alpha_s / 4\pi)\Upsilon^{(1)}$, where the pieces $\Upsilon^{(0)}$ are the same as those given in Section \ref{sec:bi-local-one-loop-finite-amps}. We determine the RI-$\overline{\text{MS}}$ matching factors from the coefficients $\Upsilon$ in the same way as was described above in Section ~\ref{sec:O_LL-one-loop-matching-factors}.

Because we evaluate a subset of our master integrals numerically using PySecDec \cite{pysecdec}, they have an inherent uncertainty associated with the quasi-Monte Carlo integration. We find absolute integration errors far below the precision reported in our tables  and we neglect them in what follows. Since the factors $\Upsilon^{(1)}$ are numerical results which are a function of $\overline{m}_c$, $\mu_{\overline{\text{MS}}}$ and $\mu_{\text{RI}}$, we don't report their values here. Instead, we will report the values for the matching factors at one- and two-loop order in the following section. 

\subsubsection{Checks of our calculation}
\label{sec:checks}
We have made several checks of our work to ensure its correctness. First of all, we have confirmed the cancellation of non-local divergences in our amplitudes after the insertion of the one-loop counterterms needed to cancel subdivergences. We find the expected GIM cancellation of the dimension-eight poles that include factors of the external momenta, leaving only poles proportional to the square of the  renormalized charm mass. Secondly, we have confirmed the $\xi$ independence of all of our counterterms, after accounting for the quark wavefunction renormalization factor. The coefficients of our $(1/\epsilon^2)$ poles correspond to the values predicted by the renormalization group from the one-loop poles, see Chapter 7 of Ref. \cite{Collins:1984xc}. Furthermore, we have checked that the values of our counterterms reproduce the anomalous dimension matrices published earlier in Refs.~\cite{Brod:2010mj, Herrlich:1996vf}. 

In order to verify the finite parts of our amplitudes, we made the following checks. In Ref. \cite{Brod:2010mj}, the same diagrams are evaluated with vanishing external momenta and with the $\lambda_u$ elimination instead of the  $\lambda_c$ elimination used here. As mentioned earlier, this changes the internal quark flavour combination for the two current-current operator insertions to be $(uu - uc)$ rather than our $(cc - uc)$. At vanishing external momenta, the $uu$ integral is scaleless and vanishes in dimensional regularization. Therefore, by comparing just our $uc$ piece we can check our results. To do this, we reevaluated all of our diagrams at vanishing external momenta in the GJK evanescent scheme, then converted to the BG evanescent scheme. At this kinematic point all of the integrals are known in closed form and we reproduced their results exactly. In order to check our finite momentum results, where some of the integrals must be evaluated numerically, we used a large $m_c$ expansion: as $\bar m_c^2 / \muRI^2 \rightarrow \infty$, we should reproduce the $q = 0$ results at the same values of $\bar m_c^2$ and $\muMS$. Our code which evaluates the $cc$ flavour structure is the same as that which evaluates $uc$ with just a mass changed in one of the fermion propagators. We tested this up to $\bar m_c^2 / \muRI^2 = 150$ and found agreement with the $q = 0$ results within the numerical integration error. For the diagrams with one current-current operator and one QCD penguin operator, the flavour combination $(cc - uu)$ is the same save a minus sign. The QCD penguin operators used in Ref. \cite{Brod:2010mj} are 
\begin{equation}
\begin{aligned}
    Q_3^{BG} & = (\bar{s} \gamma_\mu d)_{V - A} \otimes \sum_{q = u,d,s,c} (\bar{q} \gamma^\mu q)_V  \\
    Q_4^{BG} & = \sum_{a = 1}^8 (\bar{s} \gamma_\mu T^a d)_{V - A} \otimes \sum_{q = u,d,s,c} (\bar{q} \gamma^\mu T^a q)_V ,
\end{aligned}
\end{equation}
where $T^a$ are the SU(3) generators. Therefore, we can follow the same strategy as above to check our results after making the change of basis
 \begin{equation}
  \begin{aligned}
      Q_3^{BG} & = \tfrac{1}{4}\big(Q_3 + Q_5\big), \\
      Q_4^{BG} & = \tfrac{1}{8}\big(Q_4 + Q_6\big) - \frac{1}{8 N_c}\big(Q_3 + Q_5\big),
  \end{aligned}
\end{equation}
where the QCD penguin operators $Q_{3,...,6}$ are our QCD penguin operators defined in Eq.~\eqref{eq:Q-QCD-peng} and we used $\sum_a T^a_{ij}T^a_{kl} = \frac12 \delta_{il} \delta_{kj} - \frac{1}{2N_c} \delta_{ij}\delta_{kl}$ in the second line.

\subsection{Collected results}
\begin{table}[t]
  \centering
  \begin{tabular}{lccc}
  \hline\hline                                                                                                                                                        
  $\mu_{\overline{\text{MS}}}$ [GeV] & 2.643515 & 1.982637 & 1.321758\\ \hline
  $\alpha_s^{(4)}(\mu_{\overline{\text{MS}}})$ & 0.2688(50) & 0.3057(66) & 0.3826(110)\\
  $\bar m_c(\mu_{\overline{\text{MS}}})$ [GeV] & 1.1376(8) & 1.2269(35) & 1.4073(113)\\                                                                                                    
  $\hat m^2=\bar m_c^2(\mu_{\overline{\text{MS}}})/\mu_{\overline{\text{MS}}}^2$ & 0.1852(3) & 0.3829(22) & 1.1336(181)\\
  \hline\hline
  \end{tabular}
  \caption{Inputs at the three scales $\mu_{\overline{\rm MS}}=\mu_{\rm RI}$ from Ref. \cite{Yikai_thesis}. The superscript $\alpha_s^{(n)}$ on the strong coupling denotes the number of active quark flavours. The charm mass $\bar
  m_c(3\,{\rm GeV})=1.1047$~GeV and the coupling $\alpha_s^{(5)}(M_Z)=0.1184\pm0.0009$ are evolved to $\mu_{\overline{\text{MS}}}$ with four-loop running in the four-flavour theory, with the
  $b$ quark decoupled at $\mu=m_b=4.19$~GeV. The uncertainties are the propagated $\alpha_s(M_Z)$ error.}
  \label{tab:inputs}
  \end{table}
 In this section we present the values for the matching factors $Y_{ij}$ at one- and two-loop order at several values of $\mu_{\overline{\text{MS}}}$ and $\mu_{\text{RI}}$. We compare the relative size of these terms to examine the convergence of the perturbative series. We report these factors at the values of $\overline{m}_c$, $\mu_{\overline{\text{MS}}}$ and $\mu_{\text{RI}}$ which are being used by the RBC/UKQCD collaboration in their physical light quark mass calculation of $\Delta M_K$ and the long-distance contribution to $\epsilon_K$, see Refs. \cite{Huo:2025bhq, Yikai_thesis}. We have included these values in Table \ref{tab:inputs}. The strong coupling is obtained from $\alpha_s^{(5)}(M_Z)=0.1184\pm0.0009$ by four-loop running to $\mu=m_b=4.19$ GeV, three-loop decoupling of the $b$ quark, and four-loop running in the four-flavour theory to $\mu_{\overline{\text{MS}}}$, using RunDec~\cite{Rundec}. The charm mass $\bar m_c(3{\rm GeV})=1.1047$ GeV is evolved to $\mu_{\overline{\text{MS}}}$ with the same  four-flavour running.

As discussed in Sections \ref{sec:renorm-bi-local-description} and \ref{sec:convert-BG-evanescent-basis}, when we calculate the matching factors $Y_{ij}$ we must include the $O(\alpha_s)$ RI/SMOM-$\overline{\text{MS}}$ matching factors from the $\Delta S = 2$ operator $Q_{LL}$ and the scheme conversion to the BG basis used in Ref.~\cite{Brod:2010mj}. Our results are reported as
\begin{equation}
    Y^{\overline{\text{MS}}}_{ij} = Y^{(0)} + \frac{\alpha_s}{ 4\pi} \, Y^{(1)}
\end{equation}
\begin{equation}
    Y^{(1)} = Y'^{(1)} - Y^{(0)} \cdot \Delta a^{RI \rightarrow \overline{\text{MS}}}_{\Delta S = 2} + \Delta^{\text{BG}}.
\label{eq:Y-prime-to-Y}
\end{equation}
In this equation, $Y^{(1)}$ is the $O(\alpha_s)$ contribution to the bi-local operator and $Y^{(0)}$ is the $O(\alpha_s^0)$ one-loop contribution given in Eq.~\eqref{eq:bi-local-RI-projection} using the results in Table \ref{table:bi-local-one-loop-matching-factors}. The term $\Delta^{\text{BG}}$ is the evanescent scheme conversion factor from GJK to BG, and $\Delta a^{RI \rightarrow \overline{\text{MS}}}_{\Delta S = 2}$ is the $O(\alpha_s)$ RI-$\overline{\text{MS}}$ matching factor without the quark wavefunction conversion factor given in the third row of Table \ref{tab:O_LL-Delta-r-factor}. The term $Y'$ is the result of applying the RI projectors to our $O(\alpha_s)$ amplitude. $Y$ is the $O(\alpha_s)$ result which should actually be used in a lattice QCD calculation. 

In Table \ref{tab:Yprime}, we give our results for the factors $Y'^{(1)}$ evaluated in the GJK evanescent scheme at several values of $\mu_{\overline{\text{MS}}}$. Because some of our master integrals are evaluated numerically, these are reported numerically. We give separately the numerical coefficient of the gauge-fixing parameter $\xi$. In Table \ref{tab:Yalpha} we give our results for $Y^{(0)}$ and $Y^{(1)}$, evaluated in Landau gauge at several values of $\mu_{\overline{\text{MS}}}$. In these tables we see that many of the elements of $(\alpha_s / 4\pi) Y^{(1)}$ are of the same size as the one-loop values $Y^{(0)}$. However, the meaningful quantity is not the individual elements of $Y_{ij}$ themselves, but rather the quantity $Y$ contracted with the Wilson coefficients: $\sum_{ij} C_i C_j Y_{ij}$. We give these values in Table \ref{tab:YCC}, using values for the Wilson coefficients given in Table \ref{tab:Ci}. These Wilson coefficients come from Ref.~\cite{Buchalla:1995vs}, evolved to the relevant energy scales in Ref. \cite{Yikai_thesis}. Because $C_1$ and $C_2$ have opposite signs, there is a large cancellation between the large terms in $Y^{(1)}$. In this combination we see that the two-loop contributions in the parameter ranges $\muMS \leq \muRI$ take their expected size of an $O(10\%)$ correction to the one-loop result. In the ranges $\muMS > \muRI$, we observe that the NLO corrections grow to become large relative to the LO pieces, even becoming $241\%$ of LO piece for $\muMS = 2.644$ GeV and $\muRI = 1.322$ GeV. At this point the term proportional to $\ln(\muMS/\muRI)$ in the LO piece cancels a large portion of the constant term. For the points $\muMS =\muRI$ typically used in lattice QCD calculations, the correction is $-13\%$ and $+17\%$ which is of the expected size. 

Our results are to be used as follows. The numerical results for $\sum_{i,j} C_i C_j Y_{ij}$ given in Table \ref{tab:YCC} should be used in Eq.~\eqref{eq:Yij-deliverable} as a shifted Wilson coefficient for the $\Delta S = 2$ operator. These are given in units of $\bar m^2_c (\muMS) / 4\pi^2$ for values of $\bar m_c(\muMS)$ given in Table \ref{tab:inputs}. For consistency with the $\Delta S = 2$ BG evanescent scheme we used here, the Wilson coefficient $C_{\Delta S = 2}^{\MSbar}$ and the lattice to $\MSbar$ conversion factor $Z_{LL}^{\text{lat} \rightarrow \MSbar}$ should be evaluated in the BG scheme as in Refs.\cite{Gorbahn:2024qpe, Brod:2010mj}.

\begin{table}[t]\centering
  \begin{tabular}{l|rr|rr|rr}\hline\hline
   & \multicolumn{2}{c|}{$\mu_{\overline{\text{MS}}}=2.644$ GeV} & \multicolumn{2}{c|}{$\mu_{\overline{\text{MS}}}=1.983$ GeV} & \multicolumn{2}{c}{$\mu_{\overline{\text{MS}}}=1.322$ GeV}\\
  $(i,j)$ & $Y^{\prime(1)}_{ij}$ & $Y^{\prime(1)}_{ij,\xi}$ & $Y^{\prime(1)}_{ij}$ & $Y^{\prime(1)}_{ij,\xi}$ & $Y^{\prime(1)}_{ij}$ & $Y^{\prime(1)}_{ij,\xi}$\\ \hline
  (1,1) & -7.959 & 1.128 & -5.516 & 2.503 & -19.592 & 5.107\\
  (1,2),(2,1) & -6.368 & 0.376 & -11.382 & 0.834 & -24.736 & 1.702\\
  (1,3) & -10.628 & 1.348 & -2.916 & 3.654 & -17.038 & 8.199\\
  (1,4) & 24.483 & 0.449 & 16.939 & 1.218 & -3.911 & 2.733\\
  (1,5) & -2.298 & 0.318 & -11.409 & -2.401 & 15.626 & -7.498\\
  (1,6) & 36.273 & 0.106 & 38.152 & -0.800 & 60.428 & -2.499\\
  (2,2) & -8.902 & 1.114 & -10.018 & 1.429 & -16.324 & 2.054\\
  (2,3) & -6.541 & 0.449 & -9.128 & 1.218 & -21.841 & 2.733\\
  (2,4) & 1.410 & 2.019 & 6.138 & 2.552 & 7.013 & 3.641\\
  (2,5) & -3.956 & 0.106 & 6.569 & -0.800 & 37.090 & -2.499\\
  (2,6) & 26.014 & -0.862 & 26.118 & -1.545 & 46.413 & -2.907\\
  \hline\hline\end{tabular}
  \caption{The $(\gamma_\mu)$-projected two-loop bi-local amplitude $Y^{\prime(1)}_{ij}$ at $\mu_{\overline{\rm MS}}=\mu_{\rm RI}$, in units of
  $(\alpha_s/4\pi)\,\bar m_c(\mu_{\overline{\text{MS}}})^2/4\pi^2$, in the GJK evanescent scheme. Here $Y^{\prime(1)}_{ij,\xi}$ is the coefficient of the gauge
  parameter $\xi$. Inputs as in Table~\ref{tab:inputs}.}
  \label{tab:Yprime}\end{table}

 \begin{table}[t]\centering
    \begin{tabular}{l|rr|rr|rr}\hline\hline
     & \multicolumn{2}{c|}{$\mu_{\overline{\text{MS}}}=2.644$ GeV} & \multicolumn{2}{c|}{$\mu_{\overline{\text{MS}}}=1.983$ GeV} &
  \multicolumn{2}{c}{$\mu_{\overline{\text{MS}}}=1.322$ GeV}\\
    $(i,j)$ & $Y^{(0)}_{ij}$ & $\frac{\alpha_s}{4\pi}Y^{(1)}_{ij}$ & $Y^{(0)}_{ij}$ & $\frac{\alpha_s}{4\pi}Y^{(1)}_{ij}$ & $Y^{(0)}_{ij}$ &
    $\frac{\alpha_s}{4\pi}Y^{(1)}_{ij}$\\ \hline
    (1,1) & 2.857(1) & -3.222(60) & 3.765(9) & -3.509(76) & 5.799(35) & -4.552(131)\\
    (1,2),(2,1) & 0.952(1) & -1.579(29) & 1.255(3) & -1.885(41) & 1.933(12) & -2.677(77)\\
    (1,3) & 5.259(2) & 0.260(5) & 6.745(14) & 0.641(14) & 10.150(61) & 0.821(24)\\
    (1,4) & 1.753(1) & 0.686(13) & 2.248(5) & 0.649(14) & 3.383(20) & 0.328(9)\\
    (1,5) & 0.423(5) & -0.010(1) & -2.349(24) & -0.525(11) & -7.565(84) & -0.523(15)\\
    (1,6) & 0.141(2) & 0.789(15) & -0.783(8) & 0.846(18) & -2.522(28) & 1.507(43)\\
    (2,2) & 0.952(1) & -0.612(11) & 1.255(3) & -0.692(15) & 1.933(12) & -0.968(28)\\
    (2,3) & 1.753(1) & 0.023(1) & 2.248(5) & 0.015(1) & 3.383(20) & -0.218(6)\\
    (2,4) & 1.753(1) & 0.193(4) & 2.248(5) & 0.387(8) & 3.383(20) & 0.660(19)\\
    (2,5) & 0.141(2) & -0.072(1) & -0.783(8) & 0.077(2) & -2.522(28) & 0.796(23)\\
    (2,6) & 0.141(2) & 0.570(11) & -0.783(8) & 0.553(12) & -2.522(28) & 1.080(31)\\
    \hline\hline\end{tabular}
    \caption{The two terms of $Y^{\overline{\rm MS}}_{ij}=Y^{(0)}_{ij}+(\alpha_s/4\pi)Y^{(1)}_{ij}$ with $\alpha_s(\mu_{\overline{\text{MS}}})/4\pi$ at
    $\mu_{\overline{\text{MS}}}=2.644$, $1.983$, $1.322$ GeV, both in units of $\bar m_c(\mu_{\overline{\text{MS}}})^2/4\pi^2$ and in Landau gauge. The
    uncertainties on $Y^{(0)}$ are from the uncertainty on the $\overline{\text{MS}}$-renormalized charm quark mass, which in turn results from the uncertainty on the strong coupling entering the  mass running. Those on $Y^{(1)}$ are from the charm quark mass and strong coupling given in Table \ref{tab:inputs}.}
    \label{tab:Yalpha}
  \end{table}

   \begin{table}[t]\centering
  \begin{tabular}{cc|rrr|r}\hline\hline
  $\mu_{\overline{\text{MS}}}$ [GeV] & $\mu_{{RI}}$ [GeV] & $\sum C_iC_jY^{(0)}_{ij}$ & $\frac{\alpha_s}{4\pi}\sum C_iC_jY^{(1)}_{ij}$ & $\sum C_iC_jY^{\overline{\rm MS}}_{ij}$ & NLO/LO\\ \hline
  \textbf{1.322} & \textbf{1.322} & \textbf{1.869(12)} & $\mathbf{+0.325(9)}$ & \textbf{2.195(22)} & $\mathbf{+17\%}$\\                                               
  1.322 & 1.983 & 2.121(10) & $+0.492(14)$ & 2.613(24) & $+23\%$\\
  1.322 & 2.644 & 2.389(8) & $+0.607(17)$ & 2.996(25) & $+25\%$\\
  1.983 & 1.322 & 0.794(4) & $-0.125(3)$ & 0.670(1) & $-16\%$\\                                                                                                       
  \textbf{1.983} & \textbf{1.983} & \textbf{1.075(3)} & $\mathbf{+0.048(1)}$ & \textbf{1.123(4)} & $\mathbf{+4\%}$\\
  1.983 & 2.644 & 1.359(2) & $+0.186(4)$ & 1.545(6) & $+14\%$\\
  2.644 & 1.322 & 0.171(1) & $-0.414(8)$ & $-0.243(7)$ & $-241\%$\\
  2.644 & 1.983 & 0.472(1) & $-0.243(5)$ & 0.230(4) & $-51\%$\\
  \textbf{2.644} & \textbf{2.644} & \textbf{0.768(1)} & $\mathbf{-0.098(2)}$ & \textbf{0.670(1)} & $\mathbf{-13\%}$\\
  \hline\hline\end{tabular}
  \caption{Matching factors contracted with the Wilson coefficients $C_i^{\overline{\rm MS}}(\mu_{\overline{\text{MS}}})$ of Table~\ref{tab:Ci} for the nine $(\mu_{\overline{\text{MS}}},\mu_{{RI}})$ combinations of Ref.~\cite{Yikai_thesis} in units $\bar m_c(\mu_{\overline{\text{MS}}})^2/4\pi^2$ and in Landau gauge. The factors of $\alpha_s(\mu_{\overline{\text{MS}}})/4\pi$ have been included in the $O(\alpha_s)$ column. Bold rows: $\mu_{\overline{\text{MS}}}=\mu_{{RI}}$. Uncertainties  are from
  the uncertainty on the strong coupling.}
  \label{tab:YCC}\end{table}

 \begin{table}[h]
  \centering
  \begin{tabular}{l|rrr}
  \hline\hline
   & \multicolumn{3}{c}{$C_i^{\overline{\rm MS}}(\mu_{\overline{\text{MS}}})$}\\
  Operator & $\mu_{\overline{\text{MS}}}=2.644$ GeV & $1.983$ GeV & $1.322$ GeV\\ \hline
  $Q_1$ & $-0.256384$ & $-0.306553$ & $-0.397030$\\
  $Q_2$ & $ 1.116460$ & $ 1.144824$ & $ 1.199741$\\
  $Q_3$ & $ 0.017565$ & $ 0.021340$ & $ 0.028811$\\
  $Q_4$ & $-0.043631$ & $-0.051116$ & $-0.065182$\\
  $Q_5$ & $ 0.009635$ & $ 0.010204$ & $ 0.010234$\\
  $Q_6$ & $-0.055235$ & $-0.067818$ & $-0.094352$\\
  \hline\hline
  \end{tabular}
  \caption{NDR-$\overline{\rm MS}$ Wilson coefficients of the $\Delta S=1$ operators at the three scales $\mu_{\overline{\text{MS}}}$, taken from Table~4.3 of Ref. \cite{Yikai_thesis}, where
  they are obtained at next-to-leading order following Ref.~\cite{Buchalla:1995vs}.}
  \label{tab:Ci}
  \end{table}

\clearpage

\section{Conclusion}
\label{sec:conclusion}
The indirect CP-violating parameter $\epsilon_K$ is an important precision test of the Standard Model. Lattice QCD calculations of the long-distance contribution to $\epsilon_K$ at physical quark masses are currently underway, but have so far suffered from an incomplete renormalization of the bi-local operator at $O(\alpha_s)$. In this paper we calculated the RI/SMOM-$\MSbar$ matching factors for the bi-local operator at $O(\alpha_s)$, which arises at two loops. These results remove the $O(\alpha_s)$ uncertainty in the renormalization of the bi-local operator in RBC/UKQCD calculations of $\epsilon_K$. We find corrections between $-13\%$ and $+17\%$ for $\muRI = \muMS$ relative to the leading order matching factors contracted with the $\Delta S = 1$ Wilson coefficients, when evaluated in Landau gauge. To do so, we apply the method which was used in Ref.~\cite{Gorbahn:2024qpe} for the $\Delta S = 2$ RI/SMOM-$\MSbar$ matching factors. We present our results for the matching factors at several values of the $\overline{\text{MS}}$-renormalized charm quark mass, $\muRI$ and $\muMS$ which are being used in the RBC/UKQCD calculation of the long-distance contribution to $\epsilon_K$.

An interesting aspect of this calculation is that all of the code was written by an agentic AI system Claude Code, under close supervision from the author who also dictated the evaluation strategy. We would emphasize that the agentic AI facilitated a much more rapid completion of this calculation than would otherwise have been possible. We find that perturbative QCD is an ideal application of agentic AI because of the large number of internal consistency checks we can apply to check intermediate results. A more detailed description of our agentic AI strategy can be found in Ref. \cite{LundstrumAgenticAI}.

\section{Acknowledgements}
We thank Norman Christ and Yikai Huo for suggesting this problem and their many helpful discussions of the lattice QCD calculation of $\epsilon_K$. We also thank Peter Boyle and Christoph Lehner for their interest and encouragement. The author is supported in part by the U.S. Department of Energy (DOE) grant No. DE-SC0011941 and in part by the Exascale Computing Project (17-SC-20-SC), a collaborative effort of the U.S. Department of Energy Office of Science and the National Nuclear Security Administration.

\bibliographystyle{JHEP}
\bibliography{biblio}

@article{Brod:2010mj,
    author = "Brod, Joachim and Gorbahn, Martin",
    title = "{$\epsilon_K$ at Next-to-Next-to-Leading Order: The Charm-Top-Quark Contribution}",
    eprint = "1007.0684",
    archivePrefix = "arXiv",
    primaryClass = "hep-ph",
    reportNumber = "TTP-10-16",
    doi = "10.1103/PhysRevD.82.094026",
    journal = "Phys. Rev. D",
    volume = "82",
    pages = "094026",
    year = "2010"
}

@article{Buchalla:1995vs,
    author = "Buchalla, Gerhard and Buras, Andrzej J. and Lautenbacher, Markus E.",
    title = "{Weak Decays beyond Leading Logarithms}",
    eprint = "hep-ph/9512380",
    archivePrefix = "arXiv",
    reportNumber = "SLAC-PUB-7009, SLAC-PUB-95-7009, MPI-PH-95-104, TUM-T31-100-95, FERMILAB-PUB-95-305-T",
    doi = "10.1103/RevModPhys.68.1125",
    journal = "Rev. Mod. Phys.",
    volume = "68",
    pages = "1125--1144",
    year = "1996"
}

@article{Bai:2023lkr,
    author = "Bai, Ziyuan and Christ, Norman H. and Karpie, Joseph M. and Sachrajda, Christopher T. and Soni, Amarjit and Wang, Bigeng",
    title = "{Long-distance contribution to $\varepsilon_K$ from lattice QCD}",
    eprint = "2309.01193",
    archivePrefix = "arXiv",
    primaryClass = "hep-lat",
    reportNumber = "JLAB-THY-23-3937",
    doi = "10.1103/PhysRevD.109.054501",
    journal = "Phys. Rev. D",
    volume = "109",
    number = "5",
    pages = "054501",
    year = "2024"
}

@article{Gorbahn:2024qpe,
    author = {Gorbahn, Martin and J{\"a}ger, Sebastian and Kvedarait{\.{e}}, Sandra},
    title = "{RI-(S)MOM to $ \overline{\textrm{MS}} $ conversion for B$_{K}$ at two-loop order}",
    eprint = "2411.19861",
    archivePrefix = "arXiv",
    primaryClass = "hep-ph",
    doi = "10.1007/JHEP09(2025)011",
    journal = "JHEP",
    volume = "09",
    pages = "011",
    year = "2025"
}

@article{HERRLICH199539,
title = {Evanescent operators, scheme dependences and double insertions},
journal = {Nuclear Physics B},
volume = {455},
number = {1},
pages = {39-58},
year = {1995},
issn = {0550-3213},
doi = {https://doi.org/10.1016/0550-3213(95)00474-7},
url = {https://www.sciencedirect.com/science/article/pii/0550321395004747},
author = {Stefan Herrlich and Ulrich Nierste}
}

@book{Collins:1984xc,
    author = "Collins, John C.",
    title = "{Renormalization : An Introduction to Renormalization, the Renormalization Group and the Operator-Product Expansion}",
    doi = "10.1017/9781009401807",
    isbn = "978-0-521-31177-9, 978-0-511-86739-2, 978-1-009-40180-7, 978-1-009-40176-0, 978-1-009-40179-1",
    publisher = "Cambridge University Press",
    address = "Cambridge",
    series = "Cambridge Monographs on Mathematical Physics",
    volume = "26",
    year = "1984"
}

@article{Lehner:2011fz,
    author = "Lehner, Christoph and Sturm, Christian",
    title = "{Matching factors for Delta S=1 four-quark operators in RI/SMOM schemes}",
    eprint = "1104.4948",
    archivePrefix = "arXiv",
    primaryClass = "hep-ph",
    reportNumber = "BNL-95075-2011-JA, MPP-2011-48, RBRC-897",
    doi = "10.1103/PhysRevD.84.014001",
    journal = "Phys. Rev. D",
    volume = "84",
    pages = "014001",
    year = "2011"
}

@article{Lehner-Lundstrum,
      author  = {Lehner, Christoph and Lundstrum, Erik},
      title   = {Four-flavour matching factors for {$\Delta S = 1$} four-quark operators in {RI/SMOM} schemes},
      journal = {in preparation},
      year    = {2026},
  }

@article{Sturm:2009kb,
    author = "Sturm, C. and Aoki, Y. and Christ, N. H. and Izubuchi, T. and Sachrajda, C. T. C. and Soni, A.",
    title = "{Renormalization of quark bilinear operators in a momentum-subtraction scheme with a nonexceptional subtraction point}",
    eprint = "0901.2599",
    archivePrefix = "arXiv",
    primaryClass = "hep-ph",
    reportNumber = "CU-TP-1186, KANAZAWA-09-01, RBRC-771, SHEP-09-02",
    doi = "10.1103/PhysRevD.80.014501",
    journal = "Phys. Rev. D",
    volume = "80",
    pages = "014501",
    year = "2009"
}

@article{Martinelli:1994ty,
    author = "Martinelli, G. and Pittori, C. and Sachrajda, Christopher T. and Testa, M. and Vladikas, A.",
    title = "{A general method for non-perturbative renormalization of lattice operators}",
    eprint = "hep-lat/9411010",
    archivePrefix = "arXiv",
    reportNumber = "CERN-TH-7342-94, LPTHE-ORSAY-94-52, ROME-1022-1994, SHEP-94-95-03",
    doi = "10.1016/0550-3213(95)00126-D",
    journal = "Nucl. Phys. B",
    volume = "445",
    pages = "81--108",
    year = "1995"
}

@article{Herrlich:1996vf,
    author = "Herrlich, Stefan and Nierste, Ulrich",
    title = "{The Complete {$\Delta S = 2$} Hamiltonian in the next-to-leading order}",
    eprint = "hep-ph/9604330",
    archivePrefix = "arXiv",
    reportNumber = "DESY-96-048, TUM-T31-86-96",
    doi = "10.1016/0550-3213(96)00324-0",
    journal = "Nucl. Phys. B",
    volume = "476",
    pages = "27--88",
    year = "1996"
}

@article{Huo:2025bhq,
    author = "Huo, Yikai and Christ, Norman H. and Wang, Bigeng",
    title = "{Enhanced Lattice QCD Studies on $\varepsilon_{K}$ and $\Delta M_{K}$}",
    doi = "10.22323/1.466.0239",
    journal = "PoS",
    volume = "LATTICE2024",
    pages = "239",
    year = "2025"
}

@article{Shtabovenko:2023idz,
    author = "Shtabovenko, Vladyslav and Mertig, Rolf and Orellana, Frederik",
    title = "{FeynCalc 10: Do multiloop integrals dream of computer codes?}",
    eprint = "2312.14089",
    archivePrefix = "arXiv",
    primaryClass = "hep-ph",
    reportNumber = "P3H-23-089, TTP23-056, SI-HEP-2023-27",
    doi = "10.1016/j.cpc.2024.109357",
    journal = "Comput. Phys. Commun.",
    volume = "306",
    pages = "109357",
    year = "2025"
}

@article{Shtabovenko:2020gxv,
    author = "Shtabovenko, Vladyslav and Mertig, Rolf and Orellana, Frederik",
    title = "{FeynCalc 9.3: New features and improvements}",
    eprint = "2001.04407",
    archivePrefix = "arXiv",
    primaryClass = "hep-ph",
    reportNumber = "P3H-20-002, TTP19-020, TUM-EFT 130/19",
    doi = "10.1016/j.cpc.2020.107478",
    journal = "Comput. Phys. Commun.",
    volume = "256",
    pages = "107478",
    year = "2020"
}

@article{Shtabovenko:2016sxi,
    author = "Shtabovenko, Vladyslav and Mertig, Rolf and Orellana, Frederik",
    title = "{New Developments in FeynCalc 9.0}",
    eprint = "1601.01167",
    archivePrefix = "arXiv",
    primaryClass = "hep-ph",
    reportNumber = "TUM-EFT-71-15",
    doi = "10.1016/j.cpc.2016.06.008",
    journal = "Comput. Phys. Commun.",
    volume = "207",
    pages = "432--444",
    year = "2016"
}

@article{MERTIG1991345,
title = {Feyn Calc - Computer-algebraic calculation of Feynman amplitudes},
journal = {Computer Physics Communications},
volume = {64},
number = {3},
pages = {345-359},
year = {1991},
issn = {0010-4655},
doi = {https://doi.org/10.1016/0010-4655(91)90130-D},
url = {https://www.sciencedirect.com/science/article/pii/001046559190130D},
author = {R. Mertig and M. Böhm and A. Denner}
}

@article{Lee:2012cn,
      author  = {Lee, R. N.},
      title   = {Presenting {LiteRed}: a tool for the {Loop InTEgrals REDuction}},
      journal = {arXiv:1212.2685 [hep-ph]},
      year    = {2012},
  }

@article{qgraf,
title = {Automatic Feynman Graph Generation},
journal = {Journal of Computational Physics},
volume = {105},
number = {2},
pages = {279-289},
year = {1993},
issn = {0021-9991},
doi = {https://doi.org/10.1006/jcph.1993.1074},
url = {https://www.sciencedirect.com/science/article/pii/S0021999183710740},
author = {P. Nogueira}
}

@article{tarcer,
    author = "Mertig, R. and Scharf, R.",
    title = "{TARCER: A Mathematica program for the reduction of two loop propagator integrals}",
    eprint = "hep-ph/9801383",
    archivePrefix = "arXiv",
    doi = "10.1016/S0010-4655(98)00042-3",
    journal = "Comput. Phys. Commun.",
    volume = "111",
    pages = "265--273",
    year = "1998"
}

@article{Fire6,
    author = "Smirnov, A. V. and Chukharev, F. S.",
    title = "{FIRE6: Feynman Integral REduction with modular arithmetic}",
    eprint = "1901.07808",
    archivePrefix = "arXiv",
    primaryClass = "hep-ph",
    reportNumber = "TTP19-006",
    doi = "10.1016/j.cpc.2019.106877",
    journal = "Comput. Phys. Commun.",
    volume = "247 ",
    pages = "106877",
    year = "2020"
}

@article{pysecdec,
    author = "Borowka, S. and Heinrich, G. and Jahn, S. and Jones, S. P. and Kerner, M. and Schlenk, J. and Zirke, T.",
    title = "{pySecDec: A toolbox for the numerical evaluation of multi-scale integrals}",
    eprint = "1703.09692",
    archivePrefix = "arXiv",
    primaryClass = "hep-ph",
    reportNumber = "MPP-2017-42, CERN-TH-2017-063, IPPP-17-24",
    doi = "10.1016/j.cpc.2017.09.015",
    journal = "Comput. Phys. Commun.",
    volume = "222",
    pages = "313--326",
    year = "2018"
}

@article{Rundec,
    author = "Chetyrkin, K. G. and Kuhn, Johann H. and Steinhauser, M.",
    title = "{RunDec: A Mathematica package for running and decoupling of the strong coupling and quark masses}",
    eprint = "hep-ph/0004189",
    archivePrefix = "arXiv",
    reportNumber = "DESY-00-034, TTP-00-05",
    doi = "10.1016/S0010-4655(00)00155-7",
    journal = "Comput. Phys. Commun.",
    volume = "133",
    pages = "43--65",
    year = "2000"
}

@inproceedings{LundstrumAgenticAI,
    author = "Lundstrum, Erik",
    title = "{Agentic AI for QCD perturbation theory or: how I learned to stop worrying and love the bot}",
    booktitle = "{43rd International Symposium on Lattice Field Theory}",
    eprint = "2610.37960",
    archivePrefix = "arXiv",
    primaryClass = "hep-lat",
    month = "10",
    year = "2026"
}

@phdthesis{Yikai_thesis,
    author = {Huo, Yikai},
    title  = {Enhanced Lattice Quantum Chromodynamics Studies of the Neutral Kaon Mass Difference and Indirect Charge-Parity Violation},                              
    school = {Columbia University},
    year   = {2026},
    type   = {Ph.D. thesis},                              
  }

@misc{anthropic2026claude,
  author       = {{Anthropic}},
  title        = {Claude {Code} (5.1 {Fable}) [{Large language model}]},
  year         = {2026},
  howpublished = {\url{https://claude.ai}},
  note         = {Accessed: 2026-07}
}

@article{Aoki:2007xm,
    author = "Aoki, Y. and others",
    title = "{Nonperturbative renormalization of quark bilinear operators and $B_K$ using domain wall fermions}",
    eprint = "0712.1061",
    archivePrefix = "arXiv",
    primaryClass = "hep-lat",
    reportNumber = "BNL-HET-07-11, CU-TP-1180, EDINBURGH-2007-12, KANAZAWA-07-10, RBRC-681, SHEP-07-20",
    doi = "10.1103/PhysRevD.78.054510",
    journal = "Phys. Rev. D",
    volume = "78",
    pages = "054510",
    year = "2008"
}

@article{RBC:2010qam,
    author = "Aoki, Y. and others",
    collaboration = "RBC, UKQCD",
    title = "{Continuum Limit Physics from 2+1 Flavor Domain Wall QCD}",
    eprint = "1011.0892",
    archivePrefix = "arXiv",
    primaryClass = "hep-lat",
    doi = "10.1103/PhysRevD.83.074508",
    journal = "Phys. Rev. D",
    volume = "83",
    pages = "074508",
    year = "2011"
}

@article{RBC-UKQCD:2008mhs,
    author = "Allton, C. and others",
    collaboration = "RBC-UKQCD",
    title = "{Physical Results from 2+1 Flavor Domain Wall QCD and SU(2) Chiral Perturbation Theory}",
    eprint = "0804.0473",
    archivePrefix = "arXiv",
    primaryClass = "hep-lat",
    reportNumber = "BNL-HET-08-5, EDINBURGH-2008-06, RBRC-730, SHEP-0812, BNL-HET-08/5, CU-TP-1182, Edinburgh 2008/06, KEK-TH-1232, RBRC-730,
  SHEP-0812",
    doi = "10.1103/PhysRevD.78.114509",
    journal = "Phys. Rev. D",
    volume = "78",
    pages = "114509",
    year = "2008"
}

@article{Gorbahn:2010bf,
    author = "Gorbahn, Martin and Jager, Sebastian",
    title = "{Precise MS-bar light-quark masses from lattice QCD in the RI/SMOM scheme}",
    eprint = "1004.3997",
    archivePrefix = "arXiv",
    primaryClass = "hep-ph",
    doi = "10.1103/PhysRevD.82.114001",
    journal = "Phys. Rev. D",
    volume = "82",
    pages = "114001",
    year = "2010"
}

@article{Almeida:2010ns,
    author = "Almeida, Leandro G. and Sturm, Christian",
    title = "{Two-loop matching factors for light quark masses and three-loop mass anomalous dimensions in the RI/SMOM schemes}",
    eprint = "1004.4613",
    archivePrefix = "arXiv",
    primaryClass = "hep-ph",
    doi = "10.1103/PhysRevD.82.054017",
    journal = "Phys. Rev. D",
    volume = "82",
    pages = "054017",
    year = "2010"
}

@article{Gracey:2011fb,
    author = "Gracey, J. A.",
    title = "{RI'/SMOM scheme amplitudes for quark currents at two loops}",
    eprint = "1101.5266",
    archivePrefix = "arXiv",
    primaryClass = "hep-ph",
    reportNumber = "LTH-906",
    doi = "10.1140/epjc/s10052-011-1567-8",
    journal = "Eur. Phys. J. C",
    volume = "71",
    pages = "1567",
    year = "2011"
}

\end{document}